\documentclass[lettersize,journal]{IEEEtran}
\usepackage{amsmath,amsfonts,amssymb}
\usepackage{algorithmic}
\usepackage{algorithm}
\usepackage{array}
\usepackage[caption=false,font=normalsize,labelfont=sf,textfont=sf]{subfig}
\usepackage{textcomp}
\usepackage{stfloats}
\usepackage{url}
\usepackage{verbatim}
\usepackage{graphicx}
\usepackage{cite}
\usepackage{bm}
\usepackage{setspace}
\usepackage{cuted}
\usepackage{color}
\usepackage{tabu}
\usepackage{mathrsfs}
\definecolor{mblue}{rgb}{0, 0, 0.8}
\allowdisplaybreaks[2]

\begin{document}

\title{Channel Capacity for FMCW-based Optical Wireless Integrated Sensing and Communication:\\ Asymptotic Analysis and Envelope Design}

\author{Yunfeng~Wen,
Fang~Yang,~\IEEEmembership{Senior~Member,~IEEE},
Jian~Song,~\IEEEmembership{Fellow,~IEEE},
and~Zhu~Han,~\IEEEmembership{Fellow,~IEEE}
\thanks{Part of this paper has been submitted to IEEE ICC 2026~\cite{FMCW_ISAC_ICC_2026}. This work was supported in part by National Key Research and Development Program of China under Grant 2023YFE0110600; and in part by NSF ECCS-2302469, CMMI-2222810, Toyota. Amazon and Japan Science and Technology Agency (JST) Adopting Sustainable Partnerships for Innovative Research Ecosystem (ASPIRE) JPMJAP2326. \emph{(Corresponding author: Fang~Yang.)}}
\thanks{Yunfeng~Wen is with the Department of Electronic Engineering, Tsinghua University, Beijing 100084, China (e-mail: wenyf22@mails.tsinghua.edu.cn).}
\thanks{Fang~Yang is with the Department of Electronic Engineering, Tsinghua University, Beijing 100084, China; and also with the State Key Laboratory of Widegap Semiconductor Optoelectronic Materials and Technologies, Beijing 100084, China (e-mail: fangyang@tsinghua.edu.cn).}
\thanks{Jian Song is with the Shenzhen International Graduate School, Tsinghua University, Shenzhen 518055, China (e-mail: jsong@tsinghua.edu.cn).}
\thanks{Z. Han is with the Department of Electrical and Computer Engineering at the University of Houston, Houston, TX 77004 USA, and also with the Department of Computer Science and Engineering, Kyung Hee University, Seoul, South Korea, 446-701 (e-mail: hanzhu22@gmail.com).}
}

\maketitle

\begin{abstract}
  Optical wireless integrated sensing and communication (OW-ISAC) is rapidly burgeoning as a complement and augmentation to its radio-frequency counterpart. In this paper, the channel capacity is analyzed to guide the design of a coherent OW-ISAC system based on frequency-modulated continuous wave (FMCW). Firstly, the system model of FMCW-based OW-ISAC is recast into an information-theoretic formulation, where an additional harmonic-mean constraint is imposed to ensure the sensing performance. Subsequently, both lower and upper bounds for channel capacity are derived under the imposed sensing constraint, based on which asymptotic expressions for channel capacity are presented for both low and high signal-to-noise-ratio regions. Moreover, the analysis of channel capacity provides guidance for the envelope design based on pulse amplitude modulation, whose capacity-achieving capabilities are demonstrated by numerical results. Furthermore, simulations reveal the trade-off between communication and sensing functionalities. In summary, the analysis of channel capacity under the sensing constraint provides insights into both the optimality and the practicality of OW-ISAC design.
\end{abstract}



\section{Introduction}\label{sec:introduction}
Future wireless networks are anticipated to deliver both ubiquitous communication and sensing capabilities~\cite{gonzalez_ISAC_6G_2024}. Towards this goal, International Telecommunication Union (ITU) has recognized integrated sensing and communication (ISAC) as one of the six usage scenarios for the sixth-generation (6G) wireless communication system~\cite{liu_Toward_6G_2023}. Generally, the evolution of ISAC has been driven by two overarching objectives, i.e., optimality and practicality. On the optimality side, the fundamental communication-sensing trade-off has been investigated to approach the information-theoretic limits of ISAC~\cite{xiong_fundamental_tradeoff_2023,hua_MIMOISAC_CRB_2024}. Benefiting from advances in ISAC information theory, academia and industry can pursue optimal ISAC system designs. On the practical side, substantial effort has focused on implementation aspects such as waveform design~\cite{wei_Waveform_MIMOOFDM_2024}, signal processing techniques~\cite{zhang_Overview_JRC_SP_2021}, and beamforming~\cite{zhou_ISACPT_multiuser_2024}, particularly in the radio-frequency (RF) band.

While abundant research has been conducted on RF-ISAC, the optical band offers a compelling complement. In addition to the ever-burgeoning optical fibre ISAC~\cite{he_ISAC_fibre_2023} and photonics-W-band ISAC~\cite{dong_Photonic_ISAC_TFDM_2024}, the ISAC technology can also be integrated with visible light communication (VLC) and free space optical (FSO) communication~\cite{xu_IPC_OOFDM_2025}. Specifically, the intersection between ISAC and VLC yields the concept of visible light positioning and communication (VLPC)~\cite{shi_VLSC_mCAP_2023}. In contrast to VLPC, which generally relies on cooperative targets, combining ISAC with FSO enables active sensing akin to RF-ISAC, commonly referred to as optical wireless ISAC (OW-ISAC)~\cite{wen_OISAC_magazine_2024}.

Similar to its RF counterpart, OW-ISAC research has prioritized practicality through waveform design and beamforming. On one hand, a variety of waveforms have been proposed to interoperate with existing light detection and ranging (LiDAR) modalities, e.g., optical pulse amplitude modulation (PAM)~\cite{wang_PAM_VLJCAS_2024}, pulse sequence sensing and pulse position modulation~\cite{wen_PSSPPM_ISAC_2023}, remote phase-shift LiDAR with communication~\cite{hai_Remote_PS_LiDAR_2023}, combined linear frequency modulation and continuous phase modulation~\cite{wen_FSO_LFM_CPM_2023}, optical orthogonal frequency division multiplexing (O-OFDM)~\cite{muller_OWCL_OFDM_2022,cui_OISAC_OFDM_channel_2024}, etc. On the other hand, novel beamforming techniques based on optical intelligent reflecting surface (OIRS)~\cite{wang_OIRS_ISAC_2024} and optical phased array (OPA)~\cite{li_ISAC_OPA_2024} have also been presented to replace bulky mechanical steering. However, a critical challenge with these schemes is the loss of optical phase information at the receiver~\cite{wen_DCOOFDM_ISAC_2024}, which precludes direct velocity estimation as in their RF counterparts.

To overcome this challenge, researchers have turned to the frequency-modulated continuous-wave (FMCW)-based LiDAR, which enables coherent phase detection~\cite{roriz_LiDAR_survey_2022}. A practical OW-ISAC paradigm is to modulate the intensity of an FMCW waveform with communication data~\cite{marti_FMCW_FSO_2022,zhang_PC_FMCW_ISAC_2024}. Thereby, the communication receiver can detect the communication data through a cost-effective direct detection scheme, while the sensing receiver can recover the beat signal due to its monostatic structure~\cite{chen_FMCW_communication_2019,xu_FMCW_Lidar_2020}. Consequently, the FMCW intensity and phase are exploited for communication and sensing, respectively, with mutual interference arising primarily from noise amplification during beat-signal recovery~\cite{wu_Integrating_Sensing_Communication_2022}. Nevertheless, the information-theoretic analysis of OW-ISAC is in an early stage, and the optimal design for FMCW-based OW-ISAC remains open.

Towards this goal, the channel-capacity analysis provides a principled pathway to the optimality in optical wireless communications (OWC)~\cite{chaaban_Capacity_IMDD_survey_2022}. Owing to the constraints imposed by intensity modulation and direct detection (IM/DD), the capacity of OWC under a Gaussian channel is still unknown in closed form. Instead, researchers have developed capacity bounds and asymptotics to enable a superior understanding of the performance limits~\cite{wang_tight_boung_VLC_2013}. Meanwhile, the capacity achieving distributions also provide guidance for the design of input distributions in OWC~\cite{farid_capacity_achieving_2009,chaaban_FSO_bound_2016}. However, extending these results from OWC to OW-ISAC is nontrivial due to additional sensing constraints. Although the authors in~\cite{khorasgani_OISAC_fundamental_GC2024} pioneer the information-theoretic analysis for an IM/DD-based OW-ISAC system, their results are not readily applicable to a coherent FMCW-based OW-ISAC system.

\begin{figure*}[tp]
  \centering
  \includegraphics[width=0.96\textwidth]{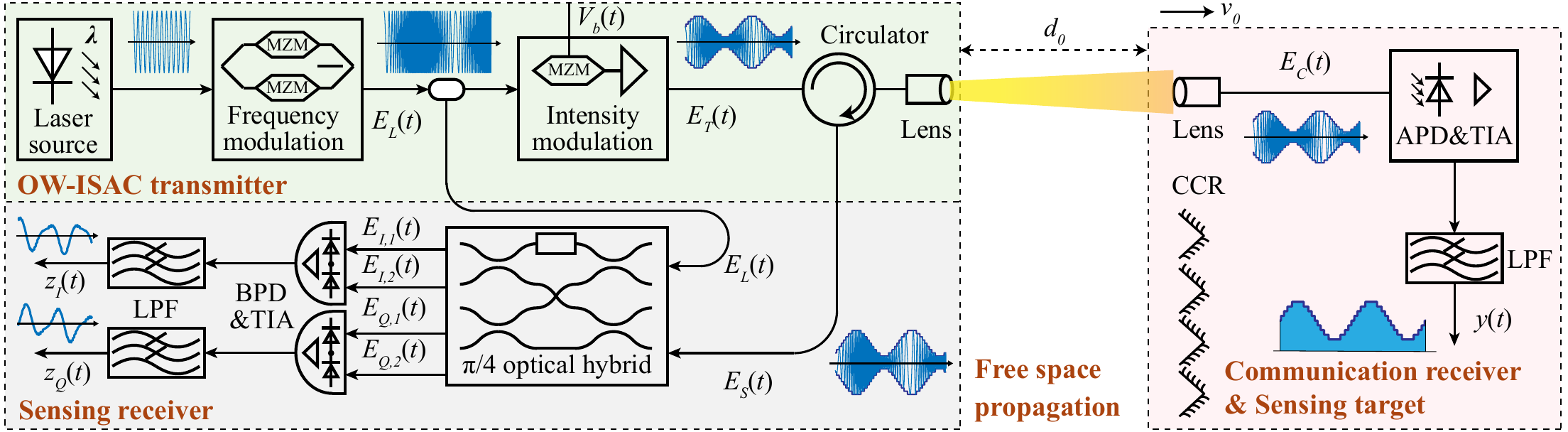}
  \caption{System model for FMCW-based OW-ISAC. The abbreviations APD, BPD, CCR, LPF, MZM, and TIA stand for avalanche photodetector, balanced photodetector, corner-cube reflector, low-pass filter, Mach-Zehnder modulator, and transimpedance amplifier, respectively.}
  \label{fig:fmcw_model}
\end{figure*}

Motivated by the potentials and challenges mentioned above, we analyze the channel capacity of an FMCW-based OW-ISAC system and investigate the optimal envelope design. Specifically, our contributions are summarized as follows:

\begin{itemize}
  \item An \textit{information-theoretic formulation} is established to capture the physical constraints of FMCW-based OW-ISAC. After presenting the system model, the direct detection for communication and the coherent detection for sensing are elaborated independently. Subsequently, the communication channel is modelled as an IM/DD Gaussian channel, while an additional harmonic-mean constraint is imposed to ensure the sensing performance, yielding a novel information-theoretic formulation.
  \item \textit{Capacity bounds} and \textit{asymptotics} are derived for FMCW-based OW-ISAC. Specifically, the entropy-power inequality (EPI) gives a lower bound of the channel capacity, for which the max-entropy distribution is also derived under the harmonic-mean constraint. In addition, leveraging the dual expression of channel capacity, auxiliary output distributions are constructed to obtain regime-specific upper bounds under different signal-to-noise-ratio (SNR) regions. Combining these upper bounds with the EPI-based lower bound yields tight asymptotic expressions for channel capacity in both regimes.
  \item PAM-based \textit{envelope design} is proposed for FMCW-based OW-ISAC under the guidance of capacity analysis. The envelope design for low SNR is given in a closed form following the discussion on the low-SNR asymptotic result. Moreover, the envelope design for high SNR is formulated as an optimization problem on the probability of each PAM level, which can be solved by the proposed extended bisection method. Numerical results indicate the capacity-achieving characteristics of the proposed PAM-based envelope design.
\end{itemize} 

The rest of this paper is organized as follows. The system model of FMCW-based OW-ISAC is introduced in Section~\ref{sec:model}. In Section~\ref{sec:capacity}, the channel model is rewritten in a information-theoretic formulation, followed by analysis on capacity bounds and asymptotics. Under the guidance of capacity analysis, the PAM-based envelope design is presented in Section~\ref{sec:envelope}. Moreover, extensive numerical results are displayed in Section~\ref{sec:simulation}, and finally the conclusion is drawn in Section~\ref{sec:conclusion}.


\section{System Model for FMCW-based OW-ISAC}\label{sec:model}
In this section, the system model of FMCW-based OW-ISAC is presented to implement concurrent communication and sensing. To lay the foundation of OW-ISAC, the generation of the FMCW-based OW-ISAC signal is first introduced in Section~\ref{subsec:isac_tx}, and the propagation of optical field is investigated in Section~\ref{subsec:propagation}. Subsequently, the direct detection for communication and the coherent detection for sensing are elaborated in Sections~\ref{subsec:comm_rx} and~\ref{subsec:sensing_rx}, respectively.

\subsection{FMCW-based OW-ISAC Signal Generation}\label{subsec:isac_tx}
As illustrated in Fig.~\ref{fig:fmcw_model}, the optical field with a single wavelength of $\lambda$ is first modulated by a dual-parallel Mach-Zehnder modulator (MZM) to generate the local reference optical field as
\begin{equation}\label{eq:local_ref}
  E_L\left(t\right)=A_L\exp\left(j2\pi\int_{0}^{t}f_L\left(\tau\right)d\tau\right),
\end{equation}

{\noindent}where $A_L$ and $f_L\left(t\right)$ denote the amplitude and instantaneous frequency of the local reference optical field, respectively. Specifically, the instantaneous frequency in a whole FMCW period is expressed as
\begin{equation}
  f_L\left(t\right)=
  \begin{cases}
    f_c+\frac{2Bt}{T_F},&\ 0\leq t\leq \frac{T_F}{2},\\
    f_c+B\left(2-\frac{2t}{T_F}\right),&\ \frac{T_F}{2}\leq t\leq T_F,\\
  \end{cases}
\end{equation}

{\noindent}where $f_c$, $B$, $T_F$ denote the optical carrier frequency, the FMCW bandwidth, and the FMCW period, respectively.

To load the communication data onto the FMCW signal, the local reference optical field in (\ref{eq:local_ref}) is fed into another MZM, which is driven by the intensity-modulation signal $V_b\left(t\right)$. According to the principle of MZM, the output optical field of the intensity modulator is written as~\cite{fu_MZM_review_2013}
\begin{equation}\label{eq:tx_mod}
  E_T\left(t\right)=A_T\cos\left(\frac{\pi V_b\left(t\right)}{2V_{\pi}}+\frac{\pi}{4}\right)\exp\left(j2\pi\int_{0}^{t}f_L\left(\tau\right)d\tau\right),\\
\end{equation}

{\noindent}where $A_T$ and $V_{\pi}$ are the amplitude of the modulated optical field and the half-wave voltage of MZM, respectively. Subsequently, the optical field in (\ref{eq:tx_mod}) is transmitted to free space by the transmitter lens.

\subsection{Propagation Model}\label{subsec:propagation}
The propagation impairments of an FMCW-based LiDAR mainly originate from geometric and misalignment losses within its maximum detection range. Without loss of generality, the proposed OW-ISAC system adopts a Gaussian beam with the transmitter lens located at the beam waist. Given the transmitted optical field of $E_T\left(t\right)$, the far-field paraxial expression of the free-space optical field is written as~\cite{goodman_Fourier_Optics}
\begin{equation}
  \begin{split}
    &\mathscr{E}\left(r,d,t\right)=\\
    &\underbrace{\frac{w_0}{w\left(d\right)}}_{\substack{\text{Geometric}\\ \text{loss}}}E_T\left(t\right)\exp\left(-jkd+\frac{j\pi}{2}\right)\underbrace{\exp\left(-\frac{r^2}{w^2\left(d\right)}\right)}_{\substack{\text{Misalignment}\\ \text{loss}}},
  \end{split}
\end{equation}

{\noindent}where $d$ and $r$ denote the axial and radial coordinates of a free space point, respectively. Besides, $k=2\pi/\lambda$ denotes the wavenumber of laser, while the beamwidth is approximated as
\begin{equation}
  w\left(d\right)=w_0\sqrt{1+\frac{d^2}{d_R^2}}\approx\frac{w_0d}{d_R},
\end{equation}

{\noindent}where $w_0$ and $d_R$ denote the waist width and the Rayleigh distance, respectively.

\begin{figure*}[!b]
  \normalsize
  \hrulefill
  \vspace*{4pt}
  \begin{subequations}
  \begin{align}
    &E_c\left(t\right)\approx\frac{A_Td_R}{d_0}\cos\left(\frac{\pi}{2V_{\pi}}V_b\left(t-\frac{d_0}{c}\right)+\frac{\pi}{4}\right)\exp\left(j2\pi\left(\int_{0}^{t-\frac{d_0}{c}}{f_L\left(\tau\right)d\tau}-\frac{f_cv_0}{c}t+\frac{1}{4}\right)-\frac{d_R^2r_c^2}{w_0^2d_0^2}\right),\label{eq:rx_field_comm}\\
    &E_s\left(t\right)\approx\frac{A_Td_R\mathfrak{R}}{d_0}\cos\left(\frac{\pi}{2V_{\pi}}V_b\left(t-\frac{2d_0}{c}\right)+\frac{\pi}{4}\right)\exp\left(j2\pi\left(\int_{0}^{t-\frac{2d_0}{c}}{f_L\left(\tau\right)d\tau}-\frac{2f_cv_0}{c}t+\frac{1}{4}\right)-\frac{d_R^2r_s^2}{w_0^2d_0^2}\right).\label{eq:rx_field_sensing}
  \end{align}
  \end{subequations}
\end{figure*}

The optical field propagates in the free space and reaches the communication receiver carried by a sensing target whose distance and velocity are $d_0$ and $v_0$, respectively. Subsequently, the communication receiver adopts a physical structure that resembles the experimental prototype in~\cite{koonen_Localization_OWC_2020} for simultaneous communication and sensing. Specifically, a lens focuses part of the optical power on an avalanche photodetector (APD). Denoting the deviation of the communication receiver from the beam center as $r_c$, its received optical field is given by~(\ref{eq:rx_field_comm}) at the bottom of this page, where $c$ denotes the velocity of light.

Meanwhile, small and compactly-mounted corner-cube reflectors (CCRs) near the receiver lens for communication generate echoes that return to the transmitter lens directly. Assuming that the deviation of the CCR from the beam center is $r_s$, the expression of the received optical field for sensing is given by~(\ref{eq:rx_field_sensing}) at the bottom of this page, where $\mathfrak{R}$ is the equivalent reflectivity of the CCR. 

\subsection{Direct Detection for Communication}\label{subsec:comm_rx}
The APD of the communication receiver responds to the magnitude of the optical field, whose output is amplified by a transimpedance amplifier (TIA). Subsequently, a low-pass filter (LPF) removes the high-frequency components, yielding the received signal for communication as
\begin{equation}\label{eq:comm_rx}
  y\left(t\right)=\mathcal{R}_c\lvert E_c\left(t\right) \rvert^2+n_c\left(t\right)\triangleq h_cx\left(t-\frac{d_0}{c}\right)+n_c\left(t\right),
\end{equation}

{\noindent}where $n_c\left(t\right)$ and $\mathcal{R}_c$ denote the post-detection noise and the responsivity of the APD-TIA pair, respectively. Without loss of generality, the post-detection noise is modelled as additive white Gaussian noise (AWGN) with a variance of $\sigma_c^2$. Besides, by reassigning the terms of $E_c\left(t\right)$ in~(\ref{eq:rx_field_comm}), the packed parameter of transmitted amplitude and channel gain for communication is expressed as
\begin{equation}\label{eq:channel_comm}
  h_c=\frac{A_T^2d_R^2\mathcal{R}_c}{d_0^2}\exp\left(-\frac{2d_R^2r_c^2}{w_0^2d_0^2}\right),
\end{equation}

{\noindent}while the real and non-negative envelope $x\left(t\right)$ carries the communication data, i.e.,
\begin{equation}\label{eq:tx_magnitude}
  x\left(t\right)=\cos^2\left(\frac{\pi V_b\left(t\right)}{2V_{\pi}}+\frac{\pi}{4}\right)=\frac{1}{2}\left(1-\sin\left(\frac{\pi V_b\left(t\right)}{V_{\pi}}\right)\right).
\end{equation}

Through the reassignment in~(\ref{eq:channel_comm}) and (\ref{eq:tx_magnitude}), the communication data is carried by a normalized envelope $x\left(t\right)$, which is independent from the propagation impairments. Furthermore, the driven signal for intensity modulation should be restricted as $V_b\left(t\right)\in[-V_\pi/2,V_\pi/2]$ to avoid ambiguity during the direct detection, which restricts the value of $x\left(t\right)$ within $[0,1]$~\cite{xu_FMCW_Lidar_2020}.

\subsection{Coherent Detection for Sensing}\label{subsec:sensing_rx}
The reflected optical field enters the receiver lens for sensing and then passes through a circulator. Subsequently, it is mixed with the local reference optical field by a $\pi/4$ optical hybrid~\cite{guan_Optical_Hybrid_2017}. Given the input of $E_s\left(t\right)$ and $E_L\left(t\right)$, the output of the $\pi/4$ optical hybrid contains the mixture of in-phase and quadrature components, i.e.,
\begin{subequations}\label{eq:rx_hybrid}
  \begin{align}
    &E_{I,1}\left(t\right)=\frac{1}{\sqrt{2}}\left(E_L\left(t\right)+E_S\left(t\right)\right),\\
    &E_{I,2}\left(t\right)=\frac{1}{\sqrt{2}}\left(E_L\left(t\right)-E_S\left(t\right)\right),\\
    &E_{Q,1}\left(t\right)=\frac{1}{\sqrt{2}}\left(E_L\left(t\right)+jE_S\left(t\right)\right),\\
    &E_{Q,2}\left(t\right)=\frac{1}{\sqrt{2}}\left(E_L\left(t\right)-jE_S\left(t\right)\right).
  \end{align}
\end{subequations}

To filter out the beat terms, two balanced photodetectors (BPDs) are utilized to detect the optical fields in~(\ref{eq:rx_hybrid}) differentially~\cite{zhang_PC_FMCW_ISAC_2024}. After the removal of high-frequency components by LPFs, the in-phase and quadrature branches of the received sensing signal are expressed as
\begin{subequations}\label{eq:rx_beat}
  \begin{align}
    \begin{split}
      &z_I\left(t\right)=\mathcal{R}_s\left(\lvert E_{I,1}\left(t\right) \rvert^2-\lvert E_{I,2}\left(t\right) \rvert^2\right)+n_{s,I}\left(t\right)\\
      &\triangleq h_sx^{\frac{1}{2}}\left(t-\frac{2d_0}{c}\right)\cos\left(2\pi f_bt+\varphi_{s,I}\right)+n_{s,I}\left(t\right),
    \end{split}\\
    \begin{split}
      &z_Q\left(t\right)=\mathcal{R}_s\left(\lvert E_{Q,1}\left(t\right) \rvert^2-\lvert E_{Q,2}\left(t\right) \rvert^2\right)+n_{s,Q}\left(t\right)\\
      &\triangleq h_sx^{\frac{1}{2}}\left(t-\frac{2d_0}{c}\right)\sin\left(2\pi f_bt+\varphi_{s,I}\right)+n_{s,Q}\left(t\right),
    \end{split}
  \end{align}
\end{subequations}

{\noindent}where we assume that the BPD-TIA pairs have an identical responsivity $\mathcal{R}_s$, and the beat frequency is calculated as
\begin{equation}\label{eq:rx_beat_freq}
  f_b=
  \begin{cases}
    \dfrac{2f_cv_0}{c}+\dfrac{4Bd_0}{cT_F},\ &t\in\left[\dfrac{2d_0}{c},\dfrac{T_F}{2}\right],\\
    \dfrac{2f_cv_0}{c}-\dfrac{4Bd_0}{cT_F},\ &t\in\left[\dfrac{T_F}{2}+\dfrac{2d_0}{c},T_F\right].
  \end{cases}
\end{equation}

{\noindent}Besides, the post-detection noise $n_{s,I}\left(t\right)$ and $n_{s,Q}\left(t\right)$ are modelled as AWGN, while $\varphi_{s,I}$ is a time-independent phase term. In addition, the packed parameter of transmitted amplitude and channel gain for sensing is expressed as
\begin{equation}
  h_s=\frac{2A_Td_R\mathfrak{R}\mathcal{R}_s}{d_0}\exp\left(-\frac{d_R^2r_s^2}{w_0^2d_0^2}\right).
\end{equation}

Since the distance and velocity to be estimated are contained in the beat frequency, the sensing receiver recovers the beat signal as
\begin{equation}\label{eq:rx_rec}
  z\left(t\right)=\frac{z_I\left(t\right)+jz_Q\left(t\right)}{\hat{x}^{\frac{1}{2}}\left(t-\frac{2d_0}{c}\right)},
\end{equation}

{\noindent}where $\hat{x}\left(t\right)$ is the estimated envelope. For simplicity, a sub-optimal but feasible estimation is given by~\cite{xu_FMCW_Lidar_2020}
\begin{equation}
  \hat{x}\left(t-\frac{2d_0}{c}\right)=\frac{z_I^2\left(t\right)+z_Q^2\left(t\right)}{h_s^2}.
\end{equation}

{\noindent}Once the beat signal is recovered, the beat frequency can be estimated by the short-time Fourier transform as
\begin{equation}
  \hat{f}_b=\arg\ \mathop{\max}\limits_{f}\Bigg| \int_{T_1}^{T_2}z\left(t\right)\exp\left(-j2\pi ft\right)dt \Bigg|,
\end{equation}

{\noindent}where the integral interval $\left[T_1,T_2\right]$ is within a single up or down ramp of FMCW. Consequently, the target distance and velocity can be obtained via~(\ref{eq:rx_beat_freq}).


\section{Channel Capacity for FMCW-based OW-ISAC}\label{sec:capacity}
In this section, the channel model is first recast into an information-theoretic formulation in Section~\ref{subsec:channel}. Then, lower and upper bounds of channel capacity are analyzed in Sections \ref{subsec:lower_bound} and \ref{subsec:upper_bound}, respectively, whose high-SNR and low-SNR asymptotics are also derived.

\subsection{Information-theoretic Formulation}\label{subsec:channel}
Supposing that $x\left(t-d_0/c\right)$ and $y\left(t\right)$ are the realizations of random variables $X$ and $Y$ for a specific time index $t$, respectively, the information-theoretic formulation for~(\ref{eq:comm_rx}) is written as
\begin{equation}\label{eq:channel_model}
  Y=X+\tilde{N}_c,
\end{equation}

{\noindent}where $\tilde{N}_c$ is the normalized AWGN with a variance of $\tilde{\sigma}_c^2=\sigma_c^2/h_c^2$. According to the definition in~(\ref{eq:tx_magnitude}), the channel for communication is a special case of the direct-detection channel with Gaussian post-detection noise.

As a significant difference between OW-ISAC and OWC, the channel input $X$ is restricted by an additional sensing constraint. However, a direct analysis on the estimation error of beat frequency is challenging due to the randomness of $X$. Instead, the error of beat signal recovery provides a more tractable metric for sensing. Specifically, if $\hat{x}\left(t\right)$ approximates $x\left(t\right)$ well in~(\ref{eq:rx_rec}), the recovered beat signal is rewritten as
\begin{equation}
  z\left(t\right)\approx\exp\left(j\left(2\pi f_bt+\varphi_{s,I}\right)\right)+\tilde{n}_s\left(t\right),
\end{equation}

{\noindent}where $\tilde{n}_s\left(t\right)$ is the normalized noise for sensing, i.e.,
\begin{equation}
  \tilde{n}_s\left(t\right)=\frac{n_{s,I}\left(t\right)+jn_{s,Q}\left(t\right)}{h_sx^{\frac{1}{2}}\left(t-\frac{2d_0}{c}\right)}.
\end{equation}

Supposing that $n_{s,I}\left(t\right)$ and $n_{s,Q}\left(t\right)$ follow an identical Gaussian distribution $\mathcal{N}\left(0,\sigma_s^2/2\right)$ independently, $\tilde{n}_s\left(t\right)$ is a complex AWGN with a variance of
\begin{equation}\label{eq:rec_mse}
  \text{var}\left(\tilde{n}_s\left(t\right)\right)=\frac{\sigma_s^2}{h_s^2}\mathbb{E}\left[\frac{1}{X}\right],
\end{equation}

{\noindent}where $\mathbb{E}\left[\cdot\right]$ means calculating the expectation with respect to all the random variables.

As a result, while the beat signal is recovered in~(\ref{eq:rx_rec}), the noise is also amplified due to the fact that $0<\hat{x}\left(t\right)\leq 1$, which may severely deteriorate the sensing performance. Since $\text{var}\left(\tilde{n}_s\left(t\right)\right)$ is an increasing function of $\mathbb{E}\left[1/X\right]$, the harmonic mean of $X$ should be larger than a threshold $1/\varsigma$ to mitigate the notorious noise amplification, and the minimum value of $X$ should also be larger than a threshold $A>0$ to avoid an ill-conditioned inversion in~(\ref{eq:rx_rec}). In addition, the principle of MZM also imposes a maximum-value constraint of $X\leq B$ with $B$ denoting a positive threshold.

Thereby, the goal of channel-capacity analysis is to obtain the optimal probability distribution function (PDF) $f_X\left(x\right)$ of $X$ to maximize the mutual information $I\left(X;Y\right)$, i.e.,
\begin{subequations}\label{eq:capacity_opt}
  \begin{align}
    &\text{(P0):} & &C=\mathop{\mathrm{\max}}\limits_{f_X}\ & &I\left(X;Y\right),&\label{eq:capacity_opt:obj}\\
    & & &\quad\quad\ \text{s.t.}\ & &\mathbb{E}\left[1/X\right]\leq\varsigma,\label{eq:capacity_opt:harmonic_constr}\\
    & & &\ & &\text{Pr}\left\{X<A\right\}=0,\label{eq:capacity_opt:min_value_constr}\\
    & & &\ & &\text{Pr}\left\{X>B\right\}=0,\label{eq:capacity_opt:max_value_constr}
  \end{align}
\end{subequations}

{\noindent}among which (\ref{eq:capacity_opt:harmonic_constr}), (\ref{eq:capacity_opt:min_value_constr}), and (\ref{eq:capacity_opt:max_value_constr}) denote harmonic-mean, minimum-value, and maximum-value constraints, respectively. Besides, the notation $\text{Pr}\left\{\cdot\right\}$ means the probability of a random event, while the minimum and maximum values are restricted by $A<B\leq 1$ according to~(\ref{eq:tx_magnitude}).

Nevertheless, the capacity of the OW-ISAC channel is not amenable to the classic methods in the RF-ISAC. Instead, the lower and upper bounds are derived based on information-theoretic fundamentals, from which asymptotic results can be obtained under both high-SNR and low-SNR approximations.

\subsection{Lower Bound with Max-entropy Distribution}\label{subsec:lower_bound}
One of the classic methods for lower bound derivation is to adopt the EPI in~\textit{Lemma 1}.

\textit{Lemma 1~\cite[Th. 17.7.3]{cover_Information_Theory}: }If $X$ and $Z$ are independent random variables with PDFs, then
\begin{equation}
  \exp\left(2h\left(X+Z\right)\right)\geq\exp\left(2h\left(X\right)\right)+\exp\left(2h\left(Z\right)\right),
\end{equation}

{\noindent}where $h\left(X\right)$ denotes the differential entropy of $X$.

\textit{Lemma 1} indicates a capacity lower bound as
\begin{equation}\label{eq:epi_bound}
  \begin{split}
    &C\geq I\left(X;Y\right)\\
    &=h\left(X+\tilde{N}_c\right)-h\left(\tilde{N}_c\right)\\
    &\geq\frac{1}{2}\log\left(\exp\left(2h\left(X\right)\right)+\exp\left(2h\left(\tilde{N}_c\right)\right)\right)-h\left(\tilde{N}_c\right)\\
    &=\frac{1}{2}\log\left(1+\frac{\exp\left(2h\left(X\right)\right)}{2\pi e\tilde{\sigma}_c^2}\right).
  \end{split}
\end{equation}

{\noindent}Thus, the derivation of lower bound is equivalent to deriving the PDF of $X$ that maximizes $h\left(X\right)$, i.e., max-entropy distribution. Towards this end, a functional optimization problem for $f_X\left(x\right)$ is formulated as
\begin{subequations}\label{eq:entropy_opt}
  \begin{align}
    &\text{(P1):} & &\mathop{\max}\limits_{f_X\left(x\right)}\ & &\int_{A}^{B}-f_X\left(x\right)\log\left(f_X\left(x\right)\right)dx,&\label{eq:entropy_opt:obj}\\
    & & &\quad\text{s.t.}\ & &\int_{A}^{B}\frac{1}{x}f_X\left(x\right)dx\leq\varsigma,&\label{eq:entropy_opt:harmonic_constr}\\
    & & &\ & &\int_{A}^{B}f_X\left(x\right)dx=1,&\label{eq:entropy_opt:normalize_constr}\\
    & & &\ & &f_X\left(x\right)\geq 0,\ x\in\left[A,B\right],&\label{eq:entropy_opt:nonnegative_constr}
  \end{align}
\end{subequations}

{\noindent}where the harmonic-mean constraint in~(\ref{eq:entropy_opt:harmonic_constr}) is equivalent to that in~(\ref{eq:capacity_opt:harmonic_constr}), while the integral limits of $A$ and $B$ correspond to (\ref{eq:capacity_opt:min_value_constr}) and (\ref{eq:capacity_opt:max_value_constr}), respectively. Besides, the PDF should also be normalized and non-negative, as depicted by~(\ref{eq:entropy_opt:normalize_constr}) and (\ref{eq:entropy_opt:nonnegative_constr}), respectively. 

As stated in~\cite[Th. 12.1.1]{cover_Information_Theory}, the optimal solution to (P1), i.e., the PDF of the max-entropy distribution, belongs to a truncated exponential family, whose expression is given by
\begin{equation}\label{eq:max_entropy_pdf}
  f_X^*\left(x\right)=
  \begin{cases}
    \exp\left(\frac{\eta^*}{x}+\mu^*-1\right),\ &A\leq x\leq B,\\
    0,\ &\text{otherwise}.
  \end{cases}
\end{equation}

{\noindent}In addition, the optimal dual variables $\eta^*$ and $\mu^*$ can be obtained under the constraints of (\ref{eq:entropy_opt:harmonic_constr}) and (\ref{eq:entropy_opt:normalize_constr}), which can be rewritten as
\begin{subequations}
  \begin{align}
    &\exp\left(\mu^*-1\right)\left(\text{Ei}\left(\frac{\eta^*}{A}\right)-\text{Ei}\left(\frac{\eta^*}{B}\right)\right)\leq\varsigma,\label{eq:capacity_opt:harmonic_constr_dual}\\
    &\exp\left(\mu^*-1\right)\mathcal{I}_h\left(A,B,\eta^*\right)=1.\label{eq:normalize_constr_dual}
  \end{align}
\end{subequations}

{\noindent}The notation $\text{Ei}\left(x\right)$ is a special function named \textit{exponential integral} and is defined as
\begin{equation}
  \text{Ei}\left(x\right)\triangleq\int_{-\infty}^{x}\frac{\exp\left(u\right)}{u}du.
\end{equation}

{\noindent}Moreover, the integral $\mathcal{I}_h\left(A,B,\eta\right)$ is defined as
\begin{equation}
  \begin{split}
    &\mathcal{I}_h\left(A,B,\eta\right)\triangleq\int_{A}^{B}\exp\left(\frac{\eta}{x}dx\right)\\
    &=B\exp\left(\frac{\eta}{B}\right)-A\exp\left(\frac{\eta}{A}\right)+\eta\text{Ei}\left(\frac{\eta}{A}\right)-\eta\text{Ei}\left(\frac{\eta}{B}\right).
  \end{split}
\end{equation}

Since (\ref{eq:capacity_opt:harmonic_constr_dual}) involves an inequality, three cases are considered as follows to achieve the optimal dual variables.

\begin{figure}[!t]
  \begin{algorithm}[H]
      \caption{Extended Bisection Method for Max-entropy Distribution}
      \setstretch{1.1}
      \label{alg:ext_bisection}
      \begin{algorithmic}[1]
          \renewcommand{\algorithmicrequire}{\textbf{Input:}}
          \renewcommand{\algorithmicensure}{\textbf{Output:}}
          \REQUIRE Threshold $\varsigma$, tolerance $\epsilon_\eta$.
          \ENSURE Optimal dual variables $\eta^*$ and $\mu^*$.
          \STATE $\eta^{\left(r\right)}\leftarrow 0$. Randomly generate $\eta^{\left(l\right)}<0$.
          \IF{$\varsigma\geq\varsigma_{\text{min}}$}
            \IF{$\varsigma\geq\varsigma_{\text{max}}$}
              \STATE $\eta^*\leftarrow 0$. \quad\quad\quad\quad\quad\quad\quad\{\textit{Case I-2}\}
            \ELSE
              \WHILE{$g_h\left(\eta^{\left(l\right)}\right)<1/\varsigma$}
                \STATE $\eta^{\left(l\right)}\leftarrow 2\eta^{\left(l\right)}$.
              \ENDWHILE
              \WHILE{$\lvert \eta^{\left(l\right)}-\eta^{\left(r\right)} \rvert\geq \epsilon_\eta$}
                \STATE $\eta^{\left(m\right)}\leftarrow\left(\eta^{\left(l\right)}+\eta^{\left(r\right)}\right)/2$.
                \IF{$\left(g_h\left(\eta^{\left(m\right)}\right)-1/\varsigma\right)\left(g_h\left(\eta^{\left(l\right)}\right)-1/\varsigma\right)<0$}
                  \STATE $\eta^{\left(r\right)}\leftarrow\eta^{\left(m\right)}$.
                \ELSE
                  \STATE $\eta^{\left(l\right)}\leftarrow\eta^{\left(m\right)}$.
                \ENDIF
              \ENDWHILE
              \STATE $\eta^*\leftarrow\left(\eta^{\left(l\right)}+\eta^{\left(r\right)}\right)/2$. \quad\{\textit{Case I-3}\}
            \ENDIF
          \ELSE
              \STATE The problem is infeasible.\quad\{\textit{Case I-1}\}
          \ENDIF
      \end{algorithmic}
  \end{algorithm}
\end{figure}

\subsubsection{Case I-1 (Feasibility Check)}
Since $\text{Pr}\left\{X>B\right\}=0$, the largest possible value for $\mathbb{E}\left[1/X\right]$ is $\varsigma_{\text{min}}\triangleq 1/B$. In consequence, the optimization problem (P1) is feasible only when $B\geq 1/\varsigma$.

\subsubsection{Case I-2 (Optimality Check)}
The maximum entropy is achieved by the uniform distribution on $\left[A,B\right]$ in the absence of the harmonic-mean constraint. If the harmonic mean of the uniform distribution subjects to the constraint, i.e.,
\begin{equation}
  \varsigma\geq\frac{\log\left(B\right)-\log\left(A\right)}{B-A}\triangleq\varsigma_{\text{max}},
\end{equation}

{\noindent}the harmonic-mean constraint is inactive, i.e., $\eta^*=0$.

\subsubsection{Case I-3 (Trade-off Scenario)}
If the uniform distribution on $\left[A,B\right]$ cannot subject to the harmonic-mean constraint, the equality in~(\ref{eq:capacity_opt:harmonic_constr_dual}) always holds, and one can divide (\ref{eq:normalize_constr_dual}) by (\ref{eq:capacity_opt:harmonic_constr_dual}) to obtain another equality for $\eta^*$ as
\begin{equation}\label{eq:auxiliary_func}
  g_h\left(\eta^*\right)=\frac{B\exp\left(\frac{\eta^*}{B}\right)-A\exp\left(\frac{\eta^*}{A}\right)}{\text{Ei}\left(\frac{\eta^*}{A}\right)-\text{Ei}\left(\frac{\eta^*}{B}\right)}+\eta^*=\frac{1}{\varsigma},
\end{equation}

{\noindent}where $g_h\left(\eta\right)$ is an auxiliary function to simplify the expression. Furthermore, the existence and uniqueness of $\eta^*$ is ensured by the following proposition.

\textit{Proposition 1: }The auxiliary function $g_h\left(\eta\right)$ is a monotone decreasing function on $(-\infty,0]$, whose limits are given by
\begin{subequations}
  \begin{align}
    &g_h\left(0\right)=\frac{1}{\varsigma_{\text{max}}}=\frac{B-A}{\log\left(B\right)-\log\left(A\right)},\\
    &\lim_{\eta\to-\infty}g_h\left(\eta\right)=\frac{1}{\varsigma_{\min}}=B.
  \end{align}
\end{subequations}

\textit{Proof: }See Appendix~\ref{sec:append_1}.

Enlightened by~\textit{Proposition 1}, an extended bisection method is proposed to obtain the optimal $\eta^*$, which is summarized in~\textbf{Algorithm~\ref{alg:ext_bisection}}. Once the optimal $\eta^*$ is achieved, the maximum differential entropy is calculated as
\begin{equation}
  \mathop{\max}\limits_{f_X\left(x\right)}\ h\left(X\right)=\log\left(\frac{\text{Ei}\left(\frac{\eta^*}{A}\right)-\text{Ei}\left(\frac{\eta^*}{B}\right)}{\varsigma}\right)-\eta^*\varsigma,
\end{equation}

{\noindent}based on which the lower bound for channel capacity can be obtained through~(\ref{eq:epi_bound}) as
\begin{equation}\label{eq:capacity_lower_bound}
  C\geq\frac{1}{2}\log\left(1+\frac{\exp\left(-2\eta^*\varsigma\right)}{2\pi e\tilde{\sigma}_c^2}\left(\frac{\text{Ei}\left(\frac{\eta^*}{A}\right)-\text{Ei}\left(\frac{\eta^*}{B}\right)}{\varsigma}\right)^2\right).
\end{equation}

\subsection{Upper Bounds and Asymptotics}\label{subsec:upper_bound}
The upper bound can be derived by a dual expression of the channel capacity in~\textit{Lemma 2}.

\textit{Lemma 2~\cite[Th. 2.1]{moser_Duality_Bounds}: }Denoting the transition probability of the channel model in (\ref{eq:channel_model}) as $f_{Y\vert X}\left(y\right)$ and the relative entropy between two probability measures as $D\left(\cdot\vert\vert\cdot\right)$, the channel capacity is upper bounded by
\begin{equation}
  C\leq\sup_{f_X}\ \mathbb{E}\left[D\left(f_{Y\vert X}\left(y\right) \vert\vert  f_Y\left(y\right)\right)\right],
\end{equation}

{\noindent}where $f_Y\left(y\right)$ denotes any feasible output distribution.

The challenge of leveraging \textit{Lemma 2} is to conceive an output distribution $f_Y\left(y\right)$ that simplifies the expectation. Towards this end, two upper bounds are derived for high-SNR and low-SNR regions, respectively.

\subsubsection{Low-SNR Upper Bound and Asymptotics}
The input constraints (\ref{eq:capacity_opt:harmonic_constr}), (\ref{eq:capacity_opt:min_value_constr}), and (\ref{eq:capacity_opt:max_value_constr}) are relaxed into a sole variance constraint. Denoting the largest possible variance and the corresponding mean of $X$ as $\sigma_X^2$ and $m_X$, respectively, the output distribution is selected as~\cite{lapidoth_Capacity_FSO_2009}
\begin{equation}\label{eq:output_pdf_low}
  f_Y\left(y\right)=\frac{1}{\sqrt{2\pi\left(\tilde{\sigma}_c^2+\sigma_X^2\right)}}\exp\left(-\frac{\left(y-m_X\right)^2}{2\tilde{\sigma}_c^2+2\sigma_X^2}\right).
\end{equation}

Thereby, the low-SNR upper bound for channel capacity is calculated as
\begin{equation}\label{eq:capacity_upper_bound_low}
  \begin{split}
    &C\leq\mathbb{E}\left[-\int_{-\infty}^{\infty}f_{Y\vert X}\left(y\right)\log\left(f_Y\left(y\right)\right)dy\right]-\frac{1}{2}\log\left(2\pi e\tilde{\sigma}_c^2\right)\\
    &=\frac{1}{2}\log\left(\frac{1}{e}\left(1+\frac{\sigma_X^2}{\tilde{\sigma}_c^2}\right)\right)+\mathbb{E}\left[\frac{\tilde{\sigma}_c^2+\left(X-m_X\right)^2}{2\tilde{\sigma}_c^2+2\sigma_X^2}\right]\\
    &\leq\frac{1}{2}\log\left(\frac{1}{e}\left(1+\frac{\sigma_X^2}{\tilde{\sigma}_c^2}\right)\right)+\frac{\tilde{\sigma}_c^2+\sigma_X^2}{2\tilde{\sigma}_c^2+2\sigma_X^2}\\
    &=\frac{1}{2}\log\left(1+\frac{\sigma_X^2}{\tilde{\sigma}_c^2}\right),
  \end{split}
\end{equation}

{\noindent}where the largest possible variance $\sigma_X^2$ and the corresponding input distribution are given by~\textit{Proposition 2}.

\textit{Proposition 2: }For a random variable $X$ that subjects to (\ref{eq:capacity_opt:harmonic_constr}), (\ref{eq:capacity_opt:min_value_constr}), and (\ref{eq:capacity_opt:max_value_constr}), its largest variance is achieved by a Bernoulli distribution, and the largest variance is given by
\begin{equation}\label{eq:var_max}
  \sigma_X^2=
  \begin{cases}
    \dfrac{\left(B-A\right)^2}{4},\ &\varsigma\geq\dfrac{1}{2}\left(\dfrac{1}{A}+\dfrac{1}{B}\right),\\
    \dfrac{B\left(B\varsigma-1\right)}{4\varsigma},\ &\varsigma<\dfrac{1}{2A},\\
    \left(\varsigma-\dfrac{1}{B}\right)\left(\dfrac{1}{A}-\varsigma\right)A^2B^2,\ &\text{otherwise}.\\
  \end{cases}
\end{equation}

\textit{Proof: }See Appendix~\ref{sec:append_2}.

\textit{Corollary 1 (Low-SNR Asymptotics): }The OW-ISAC channel is peak-constrained as indicated by~(\ref{eq:capacity_opt:max_value_constr}), and the low-SNR asymptotic capacity can be lower bounded by~\cite[Th. 2]{prelov_Weak_Asymptotic_Capacity_1993}
\begin{equation}\label{eq:capacity_lower_bound_asymptotic}
  C\geq I\left(X;Y\right)=\frac{\sigma_X^2}{2\tilde{\sigma}_c^2}+o\left(\left(B-A\right)^2\right).
\end{equation}

{\noindent}Leveraging the fact that $\log\left(1+x\right)\approx x$ when $x\to 0$, the low-SNR asymptotic expression of the channel capacity can be derived from (\ref{eq:capacity_upper_bound_low}) and (\ref{eq:capacity_lower_bound_asymptotic}) as
\begin{equation}\label{eq:capacity_asymp_low}
  \lim_{\left(B-A\right)/\tilde{\sigma}_c\to 0^+}\frac{\tilde{\sigma}_c^2C}{\left(B-A\right)^2}=\frac{\sigma_X^2}{2\left(B-A\right)^2}.
\end{equation}

\subsubsection{High-SNR Upper Bound and Asymptotics}
The output distribution of $Y$ is expected to resemble that of $X$ in the high-SNR region. To achieve this similarity asymptotically, hyper parameters $\delta>0$ and $\tilde{\eta}>0$ are introduced~\cite{lapidoth_Capacity_FSO_2009}. Thereby, $f_Y\left(y\right)$ should resemble $f_X\left(y\right)$ for $y\in\left[A-\delta,B+\delta\right]$ and have a Gaussian roll-off for $y<A-\delta$ and $y>B+\delta$, i.e.,
\begin{equation}\label{eq:output_pdf_high}
  f_Y\left(y\right)=
  \begin{cases}
    \frac{1}{\sqrt{2\pi}\tilde{\sigma}_c}\exp\left(-\frac{\left(y-A\right)^2}{2\tilde{\sigma}_c^2}\right),\ &y<A-\delta,\\
    \frac{1}{\mathcal{J}\left(\delta,\tilde{\eta}\right)}\exp\left(\frac{\tilde{\eta}}{y}\right),\ &A-\delta\leq y\leq B+\delta,\\
    \frac{1}{\sqrt{2\pi}\tilde{\sigma}_c}\exp\left(-\frac{\left(y-B\right)^2}{2\tilde{\sigma}_c^2}\right),\ &y>B+\delta,
  \end{cases}
\end{equation}

{\noindent}where $Q\left(\cdot\right)$ is the complementary cumulative distribution function of the standard Gaussian distribution, and the normalization factor $\mathcal{J}\left(\delta,\tilde{\eta}\right)$ is defined as
\begin{equation}
  \mathcal{J}\left(\delta,\tilde{\eta}\right)\triangleq\frac{\mathcal{I}_h\left(A-\delta,B+\delta,\tilde{\eta}\right)}{1-2Q\left(\frac{\delta}{\tilde{\sigma}_c}\right)}.
\end{equation}

Following the derivation in Appendix~\ref{sec:append_3}, the high-SNR upper bound for channel capacity is calculated as
\begin{equation}\label{eq:capacity_upper_bound_high}
  \begin{split}
    &C\leq\mathbb{E}\left[-\int_{-\infty}^{\infty}f_{Y\vert X}\left(y\right)\log\left(f_Y\left(y\right)\right)dy\right]-\frac{1}{2}\log\left(2\pi e\tilde{\sigma}_c^2\right)\\
    &\leq\frac{\delta}{\sqrt{2\pi}\tilde{\sigma}_c}\exp\left(-\frac{\delta^2}{2\tilde{\sigma}_c^2}\right)-\frac{1}{2}+\log\left(\frac{\mathcal{J}\left(\delta,\tilde{\eta}\right)}{\sqrt{2\pi}\tilde{\sigma}_c}\right)\\
    &\quad\cdot\left(1-Q\left(\frac{1/\varsigma^*-A+\delta}{\tilde{\sigma}_c}\right)-Q\left(\frac{B-1/\varsigma^*+\delta}{\tilde{\sigma}_c}\right)\right)\\
    &\quad+Q\left(\frac{\delta}{\tilde{\sigma}_c}\right)-\tilde{\eta}\varsigma\Biggl(1-2Q\left(\frac{B-A+2\delta}{2\tilde{\sigma}_c}\right)\\
    &\quad+\frac{\tilde{\sigma}_c}{\sqrt{2\pi}}\Biggl(\frac{1}{A-\delta}\left(1-\exp\left(-\frac{\left(B-A+\delta\right)^2}{2\tilde{\sigma}_c^2}\right)\right)\\
    &\quad-\frac{1}{B+\delta}\left(1-\exp\left(-\frac{\delta^2}{2\tilde{\sigma}_c^2}\right)\right)\Biggr)\Biggr),
  \end{split}
\end{equation}

{\noindent}where a sub-optimal but practical choice of hyper parameters in~(\ref{eq:capacity_upper_bound_high}) is given by~\cite{lapidoth_Capacity_FSO_2009}
\begin{subequations}\label{eq:hyper_param}
  \begin{align}
    \varsigma^*&=\min\left\{\varsigma,\frac{2}{A+B}\right\},\label{eq:varsigma_star}\\
    \delta&=\tilde{\sigma}_c\log\left(1+\frac{A}{2\tilde{\sigma}_c}\right),\\
    \tilde{\eta}&=\eta^*\left(1-\exp\left(-\frac{\varsigma\delta^2}{2\tilde{\sigma}_c^2}\right)\right).
  \end{align}
\end{subequations}

%

\textit{Corollary 2 (High-SNR Asymptotics): }Combining the results in~(\ref{eq:capacity_lower_bound}) and (\ref{eq:capacity_upper_bound_high}), the squeeze theorem gives the asymptotic expression of the channel capacity for high SNR as
\begin{equation}\label{eq:capacity_asymp_high}
  \begin{split}
    &\lim_{\left(B-A\right)/\tilde{\sigma}_c\to+\infty}\left\{C-\log\left(\frac{B-A}{\tilde{\sigma}_c}\right)\right\}\\
    &=-\frac{1}{2}\log\left(2\pi e\right)+\log\left(\frac{\text{Ei}\left(\frac{\eta^*}{A}\right)-\text{Ei}\left(\frac{\eta^*}{B}\right)}{\left(B-A\right)\varsigma}\right)-\eta^*\varsigma.
  \end{split}
\end{equation}


\section{Envelope Design for FMCW-based OW-ISAC}\label{sec:envelope}
Motivated by the optimality of discrete measures in conditional Gaussian channels~\cite{chaaban_Capacity_IMDD_survey_2022}, PAM is adopted as the prototype for the envelope, whose PDF is expressed as
\begin{equation}
  \hat{f}_X\left(x\right)=\sum_{m=1}^{M}a_m\delta_D\left(x-x_m\right),
\end{equation}

{\noindent}where $M$, $\delta_D\left(\cdot\right)$, $x_m$, and $a_m$ denotes the order of PAM, the Dirac-delta function, the $m$-th PAM level, and its corresponding probability, respectively.

Although the global optimal design of $\hat{f}_X\left(x\right)$ is challenging, the capacity analysis provides guidance for the PAM-based envelope design in both low-SNR and high-SNR regions, as discussed in Sections~\ref{subsec:envelope_low} and \ref{subsec:envelope_high}, respectively.

\subsection{Envelope Design for Low SNR}\label{subsec:envelope_low}
\textit{Proposition 2} and \textit{Corollary 1} indicate that $X$ should follow a Bernoulli distribution to achieve the asymptotic channel capacity in the low-SNR region, for which the PAM order should be set as $M=2$. Besides, the level $x_2=B$ is fixed following the deduction in Appendix~\ref{sec:append_2}, while the left three parameters, i.e., $x_1$, $a_1$, and $a_2=1-a_1$, are determined by the values of $E_X\left[X\right]$ and $E_X\left[1/X\right]$, i.e.,
\begin{subequations}
  \begin{align}
    &E_X\left[X\right]=a_1x_1+\left(1-a_1\right)B,\label{eq:pam_low_mean}\\
    &E_X\left[1/X\right]=a_1/x_1+\left(1-a_1\right)/B.\label{eq:pam_low_harmonic}
  \end{align}
\end{subequations}

Solving (\ref{eq:pam_low_mean}) and (\ref{eq:pam_low_harmonic}) simultaneously yields the values of $x_1$ and $a_1$ as
\begin{subequations}
  \begin{align}
    &x_1=
    \begin{cases}
      A,\ &\varsigma\geq\dfrac{1}{2A},\\
      \dfrac{1}{2\varsigma},\ &\text{otherwise},
    \end{cases}
    \\
    &a_1=
    \begin{cases}
      \dfrac{1}{2},\ &\varsigma\geq\dfrac{1}{2}\left(\dfrac{1}{A}+\dfrac{1}{B}\right),\\
      \dfrac{\varsigma-\frac{1}{B}}{2\varsigma-\frac{1}{B}},\ &0<\varsigma<\dfrac{1}{2A},\\
      \dfrac{\varsigma-\frac{1}{B}}{\frac{1}{A}-\frac{1}{B}},\ &\text{otherwise},\\
    \end{cases}
  \end{align}
\end{subequations}

{\noindent}which completes the envelope design in the low-SNR region.

\subsection{Envelope Design for High SNR}\label{subsec:envelope_high}
As stated in \textit{Corollary 2}, the lower bound in~(\ref{eq:capacity_lower_bound}) and the upper bound in~(\ref{eq:capacity_upper_bound_high}) have an identical asymptotic result in the high-SNR region. Therefore, a discrete approximation of the max-entropy distribution in~(\ref{eq:max_entropy_pdf}) is capacity-achieving as the PAM order tends to infinity. Without loss of generality, the PAM levels are uniformly distributed on $\left[A,B\right]$, i.e.,
\begin{equation}
  x_m=\frac{\left(m-1\right)B}{M-1}+\frac{\left(M-m\right)A}{M-1},\ \ 1\leq m\leq M.
\end{equation}

Subsequently, the PAM-based envelope design is equivalent to optimizing the probability $a_m$ to approximate the max-entropy distribution. Towards this goal, the optimization problem for PAM-based envelope design is formulated as
\begin{subequations}\label{eq:pam_opt}
  \begin{align}
    &\text{(P2):} & &\mathop{\max}\limits_{a_m}\ & &-\sum_{m=1}^{M}a_m\log\left(a_m\right),&\label{eq:pam_opt:obj}\\
    & & &\quad\text{s.t.}\ & &\sum_{m=1}^{M}\frac{a_m}{x_m}\leq\varsigma,&\label{eq:pam_opt:harmonic_constr}\\
    & & &\ & &\sum_{m=1}^{M}a_m=1,&\label{eq:pam_opt:normalize_constr}\\
    & & &\ & &a_m\geq 0,\ \ 1\leq m\leq M,&\label{eq:pam_opt:nonnegative_constr}
  \end{align}
\end{subequations}

{\noindent}where constraints (\ref{eq:pam_opt:harmonic_constr}), (\ref{eq:pam_opt:normalize_constr}), (\ref{eq:pam_opt:nonnegative_constr}) are equivalent to (\ref{eq:entropy_opt:harmonic_constr}), (\ref{eq:entropy_opt:normalize_constr}), and (\ref{eq:entropy_opt:nonnegative_constr}), respectively.

Since (P2) is a convex optimization problem, the Karush-Kuhn-Tucker conditions give the optimal solution as~\cite{ahn_Inverse_Source_Coding_2012}
\begin{equation}\label{eq:pam_opt_prob}
  a_m^*=\exp\left(\frac{\hat{\eta}^*}{x_m}+\hat{\mu}^*-1\right),
\end{equation}

{\noindent}where the optimal dual variables $\hat{\eta}^*$ and $\hat{\mu}^*$ can be obtained under the constraints of (\ref{eq:pam_opt:harmonic_constr}) and (\ref{eq:pam_opt:normalize_constr}), i.e.,
\begin{subequations}
  \begin{align}
    &\exp\left(\hat{\mu}^*-1\right)\sum_{m=1}^{M}\frac{1}{x_m}\exp\left(\frac{\hat{\eta}^*}{x_m}\right)\leq\varsigma,\label{eq:capacity_opt:harmonic_constr_discrete}\\
    &\exp\left(\hat{\mu}^*-1\right)\sum_{m=1}^{M}\exp\left(\frac{\hat{\eta}^*}{x_m}\right)=1.\label{eq:normalize_constr_discrete}
  \end{align}
\end{subequations}

Since (\ref{eq:capacity_opt:harmonic_constr_discrete}) involves an inequality, three cases are considered as follows to achieve the optimal dual variables.

\subsubsection{Case II-1 (Feasibility Check)}
Since $X\leq B$, the optimization problem (\ref{eq:pam_opt}) is feasible only when $B\geq 1/\varsigma$.

\subsubsection{Case II-2 (Optimality Check)}
The maximum entropy is achieved by a equal-probability distribution in the absence of the harmonic-mean constraint, i.e., $a_m=1/M$. If the harmonic mean of the equal-probability distribution subjects to the constraint, i.e.,
\begin{equation}
  \frac{1}{M}\sum_{m=1}^{M}\frac{1}{x_m}\leq\varsigma,
\end{equation}

{\noindent}the harmonic-mean constraint is inactive, i.e., $\hat{\eta}^*=0$.

\subsubsection{Case II-3 (Trade-off Scenario)}
In case the equal-probability distribution cannot subject to the harmonic-mean constraint, the equality in~(\ref{eq:capacity_opt:harmonic_constr_discrete}) always holds, and thus the equality for $\hat{\eta}^*$ is obtained through a division between (\ref{eq:normalize_constr_discrete}) and (\ref{eq:capacity_opt:harmonic_constr_discrete}), i.e.,
\begin{equation}\label{eq:opt_dual_discrete}
  \frac{\sum_{m=1}^{M}\exp\left(\frac{\hat{\eta}^*}{x_m}\right)}{\sum_{m=1}^{M}\frac{1}{x_m}\exp\left(\frac{\hat{\eta}^*}{x_m}\right)}=\frac{1}{\varsigma}.
\end{equation}

{\noindent}Given the optimal $\hat{\eta}^*$, the optimal $\hat{\mu}^*$ is then calculated as
\begin{equation}
  \hat{\mu}^*=1+\log\left(\frac{1}{\sum_{m=1}^{M}\exp\left(\frac{\hat{\eta}^*}{x_m}\right)}\right).
\end{equation}

{\noindent}The existence and uniqueness of the solution $\hat{\eta}^*$ to (\ref{eq:opt_dual_discrete}) can be proven in a similar method to that for \textit{Proposition 1}. Therefore, one can also adopt the extended bisection method to attain the optimal $\hat{\eta}^*$ on $(-\infty,0]$. 

Once the optimal dual variables $\hat{\eta}^*$ and $\hat{\mu}^*$ are achieved, the optimal probability $a_m^*$ is calculated through~(\ref{eq:pam_opt_prob}). Thereby, the cumulative distribution function (CDF) of PAM tends to that of the max-entropy distribution as the PAM order tends to infinity, which completes the envelope design for OW-ISAC in the high-SNR region.


\begin{figure}[tp]
  \centering
  \includegraphics[width=0.48\textwidth]{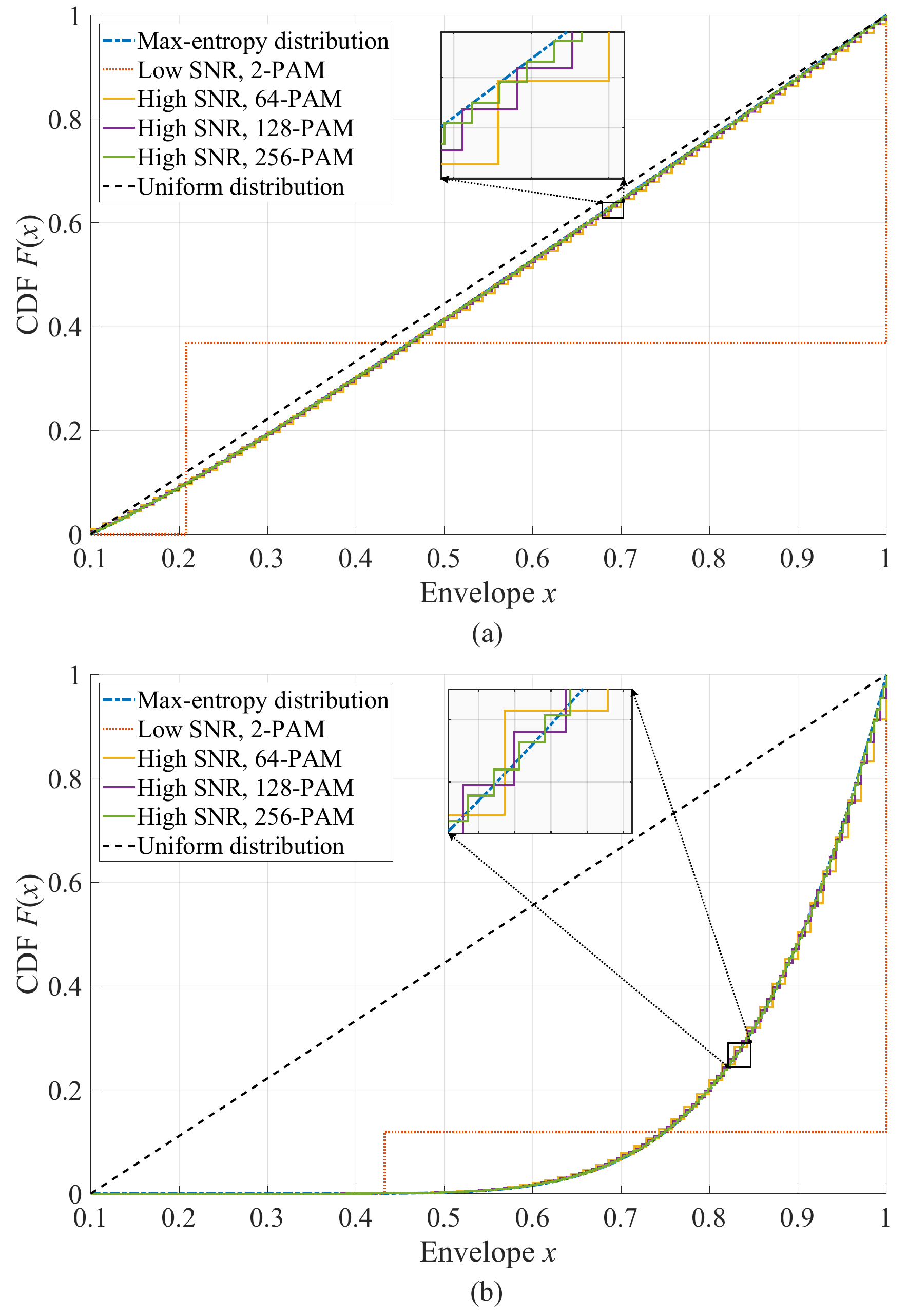}
  \caption{CDFs of the max-entropy distribution and PAM-based envelopes. (a) $10\%$ NSP with $A=0.10,B=1.00,\varsigma=1.156$. (b) $90\%$ NSP with $A=0.10,B=1.00,\varsigma=2.406$.}
  \label{fig:cdf_max_entropy}
\end{figure}

\begin{figure}[tp]
  \centering
  \includegraphics[width=0.48\textwidth]{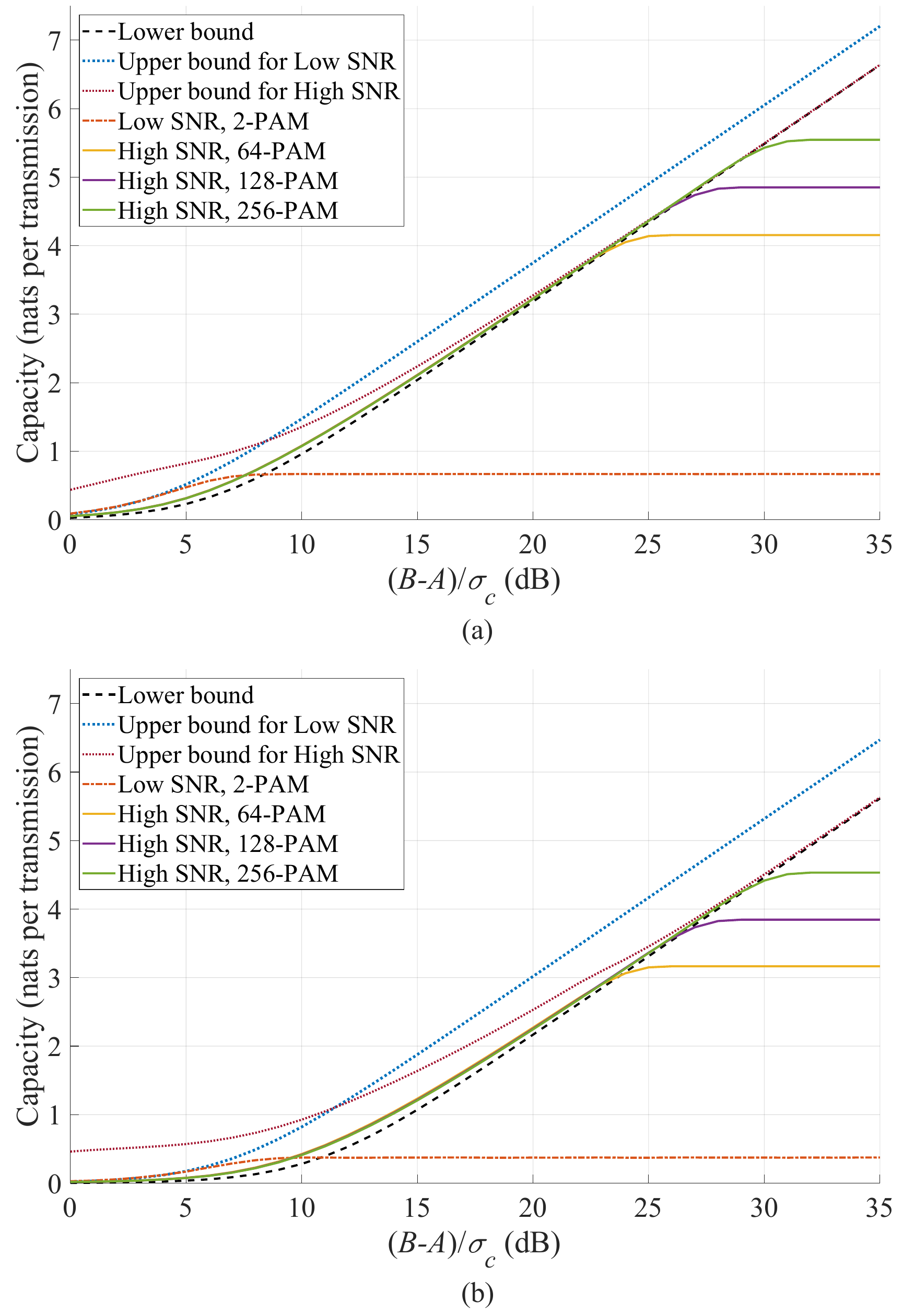}
  \caption{Capacity bounds and achievable data rates of PAM-based envelopes. (a) $10\%$ NSP with $A=0.10,B=1.00,\varsigma=1.156$. (b) $90\%$ NSP with $A=0.10,B=1.00,\varsigma=2.406$.}
  \label{fig:capacity}
\end{figure}

\section{Numerical Results}\label{sec:simulation}
This section provides numerical results to substantiate the proposed channel capacity and envelope design. Firstly, the max-entropy distribution and the corresponding channel capacity are displayed in Section~\ref{subsec:capacity_result}. Then, Monte-Carlo simulations are conducted to evaluate the sensing performance metrics in Section~\ref{subsec:sensing_result}. Subsequently, the trade-off between communication and sensing performance is elaborated in Section~\ref{subsec:tradeoff_result}.

\subsection{Max-entropy Distribution and Channel Capacity}\label{subsec:capacity_result}
The max-entropy distribution is the fundamental for the calculation of capacity bounds under different sensing constraints. Since the range of a feasible threshold $\varsigma$ depends on specific values of $A$ and $B$, a normalized sensing priority (NSP) is defined as
\begin{equation}
  \chi\triangleq\frac{\varsigma_{\max}-\varsigma}{\varsigma_{\max}-\varsigma_{\min}}\times 100\%,
\end{equation}

{\noindent}where the values of $\varsigma_{\max}$ and $\varsigma_{\min}$ are given in Section~\ref{subsec:lower_bound}. Then, tuning the NSP $\chi$ is equivalent to tuning the threshold $\varsigma$ for harmonic mean. Without loss of generality, the minimum and maximum values of the envelope are set as $A=0.10$ and $B=1.00$, respectively, while two different thresholds of $\varsigma$ are adopted to derive the CDFs in Fig.~\ref{fig:cdf_max_entropy}. 

As illustrated in Fig.~\ref{fig:cdf_max_entropy}, the optimal PDF is an increasing function of the envelope value under a trade-off scenario, i.e., $0\%<\chi<100\%$, which yields a CDF curve under that of the uniform distribution, i.e., $\chi=0\%$. Besides, an enhanced NSP forces the envelope to take large values with an increased probability, which ultimately converges to the deterministic value of $B$ when $\chi$ approaches $100\%$. Additionally, the CDF of a PAM-based envelope approximates that of the max-entropy distribution with less error as its order becomes higher. In consequence, a high-order PAM-based envelope shows similar statistical characteristics to those of the continuous max-entropy distribution.

\begin{table}[tp]
  \centering
  \renewcommand\arraystretch{1.05}
  \caption{Simulation Configurations for FMCW}
  \begin{tabular}{c|c|c}
      \hline \hline
      Parameter & Notation & Value \\
      \hline 
      Optical carrier frequency & $f_c$ & $194$ THz \\
      Velocity of light & $c$ & $3\times 10^8$ m/s \\
      FMCW bandwidth & $B$ & $5$ GHz \\
      FMCW period & $T_F$ & $10$ {\textmu}s \\
      PAM symbol per period & $N_s$ & 500 \\
      Sample rate & $R_s$ & $200$ MHz \\
      \hline \hline
  \end{tabular}
  \label{tab:sim_config}
\end{table}

As displayed in Fig.~\ref{fig:capacity}, once the max-entropy distributions and their approximations are obtained, the channel capacity can be calculated numerically following the methods in~\cite{ahn_Inverse_Source_Coding_2012}. The lower bound in~(\ref{eq:capacity_lower_bound}) and the upper bound in~(\ref{eq:capacity_upper_bound_high}) converge to the same asymptotic result in~(\ref{eq:capacity_asymp_high}) for a high SNR of $\left(B-A\right)/\sigma_c\geq 25\ \text{dB}$. In addition, the achievable data rates of PAM-based envelopes also converge to the high-SNR asymptotic capacity before reaching their maximums, which indicates their capacity-achieving capabilities. In contrast, a gap exists between the lower bound in~(\ref{eq:capacity_lower_bound}) and the upper bound in~(\ref{eq:capacity_upper_bound_low}) for a low SNR of $\left(B-A\right)/\sigma_c\leq 5\ \text{dB}$. However, the max-variance distribution given in Section~\ref{subsec:envelope_low} yields an achievable data rate that overlaps with the upper bound, which validates the optimality of 2-PAM in the low-SNR region.

\begin{figure}[tp]
  \centering
  \includegraphics[width=0.48\textwidth]{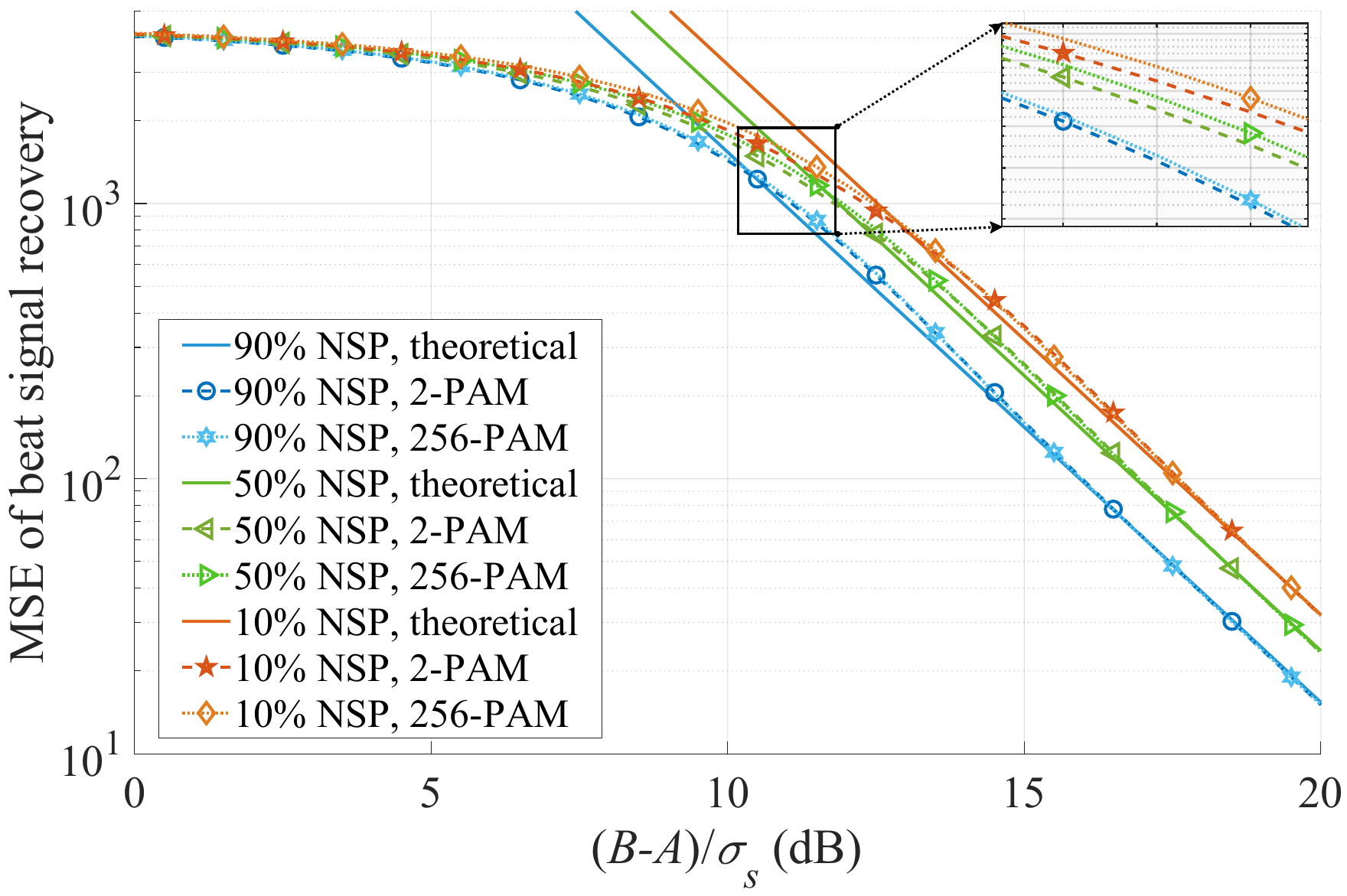}
  \caption{MSE of beat signal recovery with different NSPs.}
  \label{fig:sensing_mse}
\end{figure}

\subsection{Sensing Performance Metrics}\label{subsec:sensing_result}
To evaluate the sensing performance metrics, supplementary parameters of FMCW are established from the experimental prototype in~\cite{xu_FMCW_Lidar_2020}, as displayed in Table~\ref{tab:sim_config}. Accordingly, the mean-square error (MSE) of beat signal recovery is defined as
\begin{equation}
  \varepsilon_z^2\triangleq\sum_{n=0}^{N_s-1}\Bigg| z\left(\frac{n}{R_s}\right)-\exp\left(j\left(\frac{2\pi f_bn}{R_s}+\varphi_{s,I}\right)\right) \Bigg|,
\end{equation}

{\noindent}where $R_s$ and $N_s$ denote the sample rate and the number of PAM symbols in each FMCW period, respectively. In addition, the error of beat frequency estimation is also calculated as a conventional sensing metric in $10^4$ Monte-Carlo simulations. During each simulation, both the envelope signal $x\left(t\right)$ and the noise signal $n_s\left(t\right)$ are randomized according to their statistics, so that the ergodicity of the transmitted signal is guaranteed.

\begin{figure}[tp]
  \centering
  \includegraphics[width=0.48\textwidth]{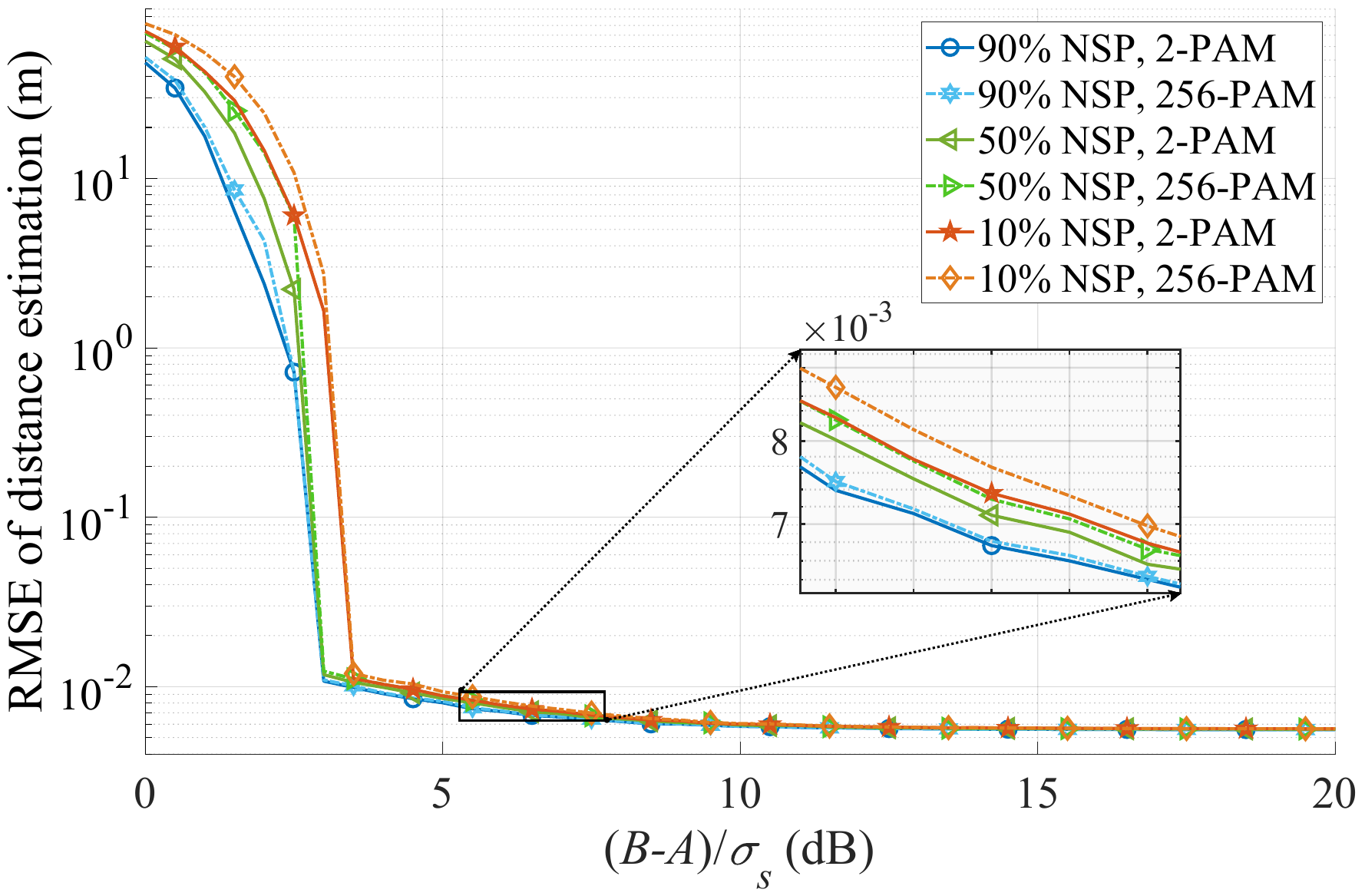}
  \caption{RMSE of target distance and velocity estimation with different NSPs.}
  \label{fig:sensing_rmse}
\end{figure}

The simulated MSE of beat signal recovery is displayed in Fig.~\ref{fig:sensing_mse}. For the low-SNR region, the phase of recovered beat signal in~(\ref{eq:rx_rec}) is randomly distributed on $[0,2\pi)$. Since the distance between a random-phase signal and a deterministic beat signal is a constant, all of the MSE curves in Fig.~\ref{fig:sensing_mse} converge to an identical value in the low-SNR region. In contrast, the recovered beat signal approximates the original beat signal well in the high-SNR region, and thus the theoretical result from~(\ref{eq:rec_mse}) serves as a reference. As illustrated in Fig.~\ref{fig:sensing_mse}, a specific NSP, i.e., a threshold $\varsigma$ for harmonic mean, leads to an identical MSE curve in the high-SNR region despite the varying order of PAM. Therefore, the MSE of beat signal recovery can be tuned by varying the threshold $\varsigma$, which rationalizes the harmonic mean as a tractable sensing metric.

\begin{figure}[tp]
  \centering
  \includegraphics[width=0.48\textwidth]{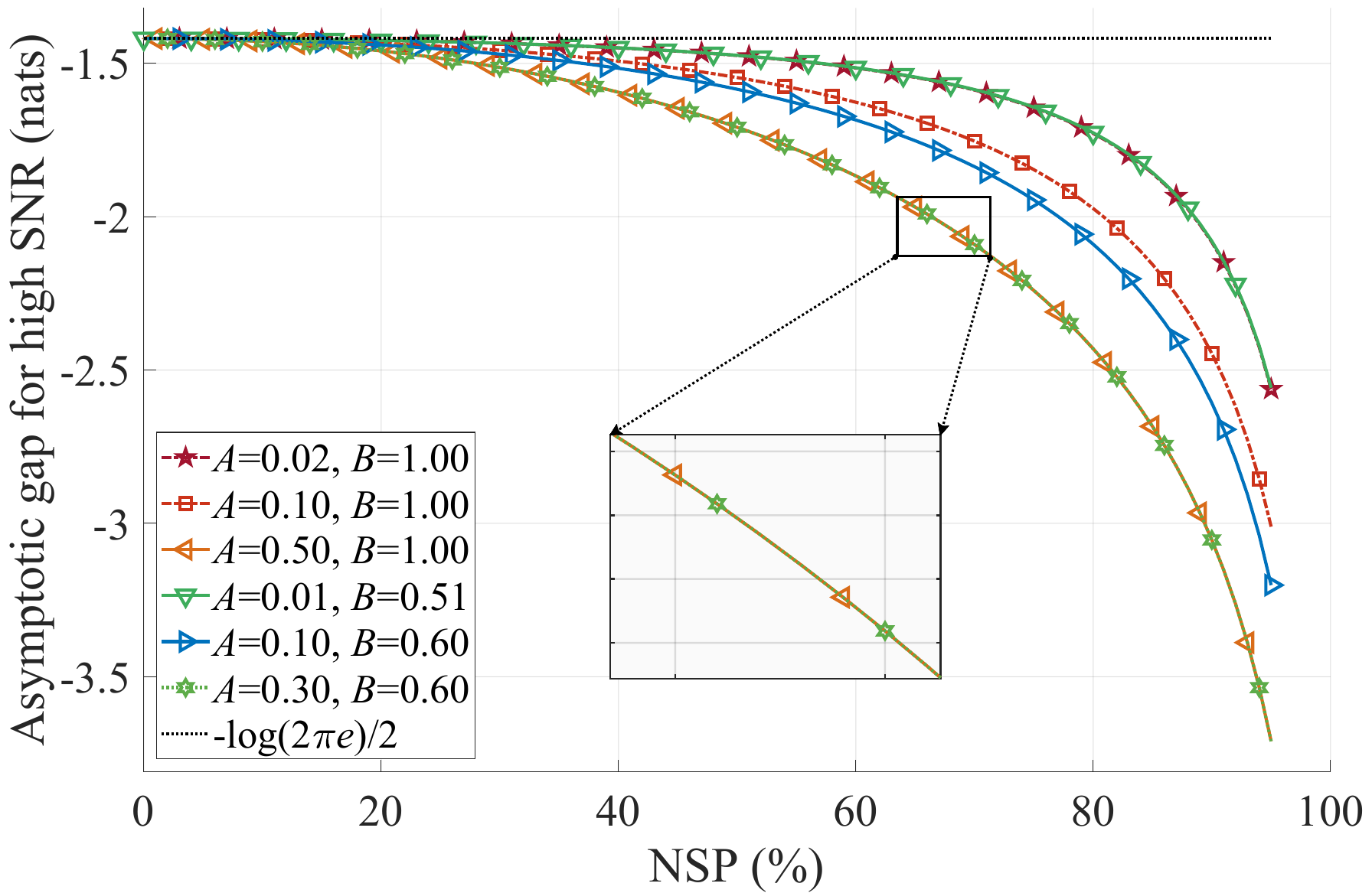}
  \caption{Asymptotic gaps versus different NSPs for high SNR.}
  \label{fig:tradeoff_high}
\end{figure}

While the harmonic-mean metric simplifies the analysis of channel capacity, the error of estimating distance is a more prevalent sensing metric for real-world systems, which is illustrated in Fig.~\ref{fig:sensing_rmse}. Firstly, the root-mean-square error (RMSE) of beat estimation, i.e., $(\mathbb{E}[\lvert\hat{f}_b-f_b\rvert^2])^{1/2}$, is obtained via numerical simulations. Then, the RMSE for distance estimation is calculated through~(\ref{eq:rx_beat_freq}) under the configurations of Table~\ref{tab:sim_config}. As shown in Fig.~\ref{fig:sensing_rmse}, the FMCW-based OW-ISAC system can achieve a range precision down to the millimeter level in the asymptotic region, which corresponds with the metric of a conventional FMCW-based LiDAR~\cite{behroozpour_lidar_system_2017}. Moreover, even though the harmonic-mean metric is not dedicated to improving the RMSE metric, the RMSE curves only have subtle differences if their harmonic means are fixed. As a result, a comparison between Fig.~\ref{fig:sensing_mse} and Fig.~\ref{fig:sensing_rmse} validates the correspondence between the tractable harmonic-mean metric and the practical RMSE metric.

\begin{figure}[tp]
  \centering
  \includegraphics[width=0.48\textwidth]{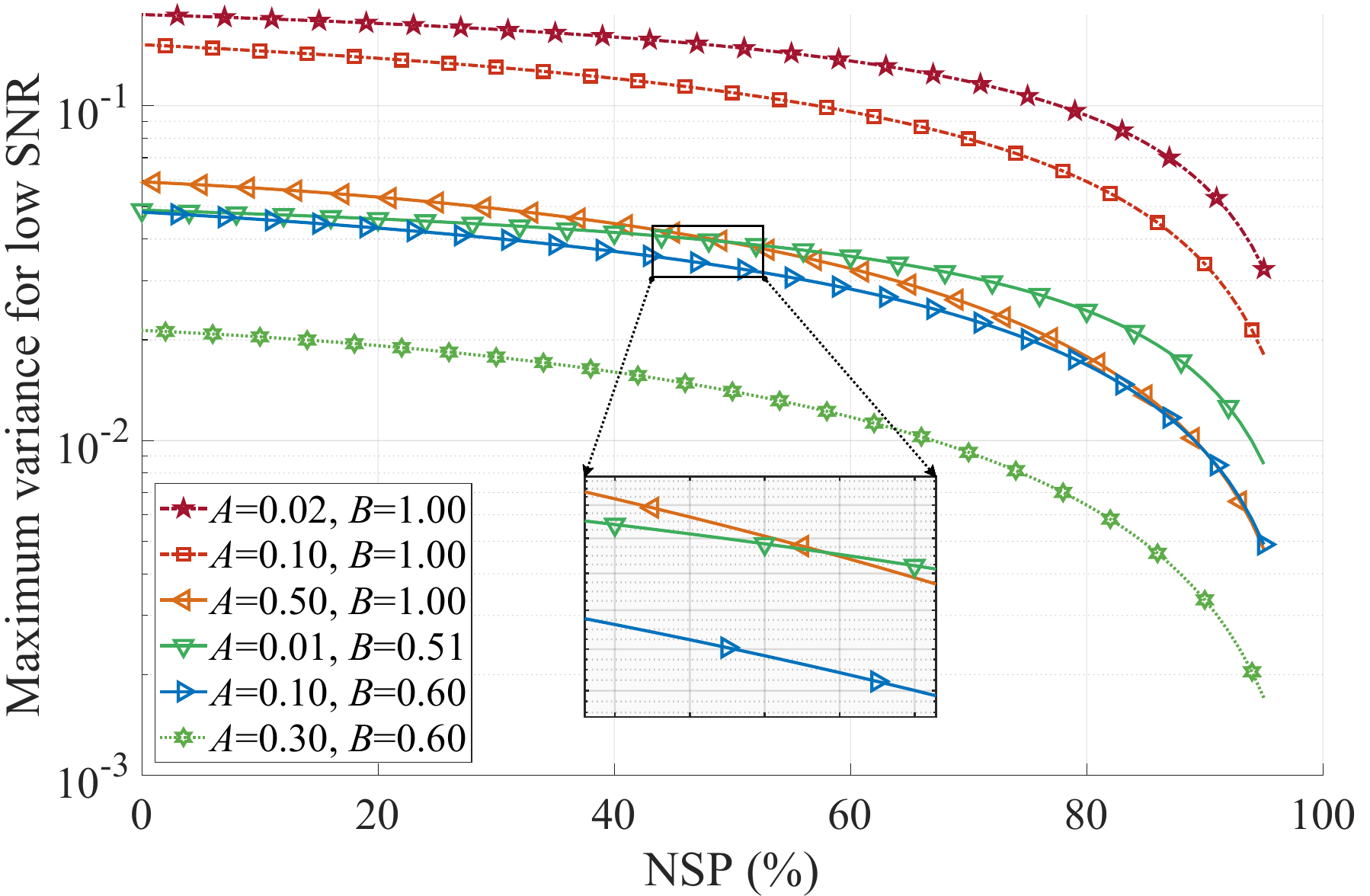}
  \caption{Maximum variances versus different NSPs for low SNR.}
  \label{fig:tradeoff_low}
\end{figure}

\subsection{Trade-off between Communication and Sensing}\label{subsec:tradeoff_result}
While the effects of tuning the threshold $\varsigma$ can be observed from Fig.~\ref{fig:cdf_max_entropy} to Fig.~\ref{fig:sensing_rmse}, this subsection provides more intuitive results for the trade-off between the NSP and the asymptotic expressions of channel capacity, where we primarily discuss the influences of hyper parameters $A$ and $B$.

As displayed in Fig.~\ref{fig:tradeoff_high}, the channel capacity for high SNR is mainly depicted by the asymptotic gap between the channel capacity and $\log\left(\left(B-A\right)/\sigma_c\right)$, i.e., the right-hand side of~(\ref{eq:capacity_asymp_high}). Specifically, the gap equals to -$\log\left(2\pi e\right)/2$ for a conventional OWC system with a sole peak-power constraint and increases as the NSP is enhanced. In addition, a critical observation from Fig.~\ref{fig:tradeoff_high} is that the position of a trade-off curve only depends on the value of $A/B$ despite the varying $A$ and $B$. The reason lies in that the solution to~(\ref{eq:auxiliary_func}) becomes $\kappa\eta^*$ if $A$, $B$, and $\varsigma$ are substituted by $\kappa A$, $\kappa B$, and $\varsigma/\kappa$, respectively, where $\kappa>0$ is an arbitrary positive number. In this case, both the NSP $\chi$ and the asymptotic gap in~(\ref{eq:capacity_asymp_high}) maintain the same, yielding an identical point in the trade-off plane.

As illustrated in Fig.~\ref{fig:tradeoff_low}, the channel capacity for low SNR is mainly delineated by the largest variance $\sigma_X^2$ in~(\ref{eq:var_max}). Generally, the low-SNR variance is dominated by the value of $\left(B-A\right)$, which is the same as that of a conventional OWC system. However, in contrast to the sole dependence of the asymptotic gap on $B/A$, the low-SNR variance is also affected by the specific values of $A$ and $B$. For instance, even if $\left(B-A\right)=0.50$ is fixed, the parameter group $(A,B)=(0.50,1.00)$ outperforms $(A,B)=(0.01,0.51)$ in sensing for an NSP of $\chi\leq 40\%$, while the latter provides a larger variance when $\chi\geq 60\%$. The reason lies in the complicated expression in~(\ref{eq:var_max}), where the largest variance does not depend on the value of $\left(B-A\right)$ solely, particularly in the case of $\varsigma<(1/A+1/B)/2$.


\section{Conclusion}\label{sec:conclusion}
In this paper, the channel capacity of an FMCW-based OW-ISAC system was investigated to achieve the optimal envelope design under the sensing constraint. An information-theoretic formulation was first established based on the system model, where the harmonic mean was adopted to delineate the sensing performance. Subsequently, the channel capacity was analyzed under the sensing constraint. Specifically, the lower bound was given by the EPI and the max-entropy distribution, while the upper bound was derived based on the dual expression of channel capacity. Moreover, the PDFs of PAM-based envelopes were designed under the guidance of the channel capacity and the corresponding asymptotic expressions, whose capacity-achieving capabilities were also demonstrated by numerical results. Furthermore, the trade-off between communication and sensing performance metrics was also revealed through various numerical simulations. Consequently, the results of channel capacity provided not only insights into the optimality of FMCW but also guidance for waveform design in OW-ISAC.


{\appendices
\section{Proof of \textit{Proposition 1}}\label{sec:append_1}
We rewrite the expression of $g_h\left(\eta\right)$ in its integral form as
\begin{equation}
  g_h\left(\eta\right)=\frac{\int_{A}^{B}\exp\left(\frac{\eta}{u}\right)du}{\int_{A}^{B}\frac{1}{u}\exp\left(\frac{\eta}{u}\right)du}.
\end{equation}

{\noindent}Then, a direct substitution of $\eta$ with $0$ yields $g_h\left(0\right)=1/\varsigma_{\max}$, while the limit $g_h\left(-\infty\right)$ can be calculated as
\begin{equation}\label{eq:aux_limit_neginf}
  \begin{split}
    &\lim_{\eta\to -\infty}g_h\left(\eta\right)\overset{\text{(a)}}{=}\lim_{\eta\to -\infty}\frac{\int_{1/B}^{1/A}\frac{1}{r^2}\exp\left(\eta r\right)dr}{\int_{1/B}^{1/A}\frac{1}{r}\exp\left(\eta r\right)dr}\\
    &\overset{\text{(b)}}{=}\lim_{\eta\to -\infty}\frac{\int_{1/B}^{1/A}\frac{1}{r}\exp\left(\eta r\right)dr}{\int_{1/B}^{1/A}\exp\left(\eta r\right)dr}\\
    &\overset{\text{(c)}}{=}\lim_{\eta\to -\infty}\frac{\eta\left(\exp\left(\frac{\eta}{A}\right)-\exp\left(\frac{\eta}{B}\right)\right)}{\frac{\eta}{A}\exp\left(\frac{\eta}{A}\right)-\frac{\eta}{B}\exp\left(\frac{\eta}{B}\right)-\exp\left(\frac{\eta}{A}\right)+\exp\left(\frac{\eta}{B}\right)}\\
    &=\lim_{\eta\to -\infty}\left(\frac{1}{B}+\frac{\frac{1}{A}-\frac{1}{B}}{1-\exp\left(\eta\left(\frac{1}{B}-\frac{1}{A}\right)\right)}-\frac{1}{\eta}\right)^{-1}\\
    &=\left(\frac{1}{B}\right)^{-1}=B,
  \end{split}
\end{equation}

{\noindent}where (\ref{eq:aux_limit_neginf}a) involves a variable substitution of $r=1/u$, while (\ref{eq:aux_limit_neginf}b) and (\ref{eq:aux_limit_neginf}c) follow the L'Hospital's rule.

In addition, the derivative of $g_h\left(\eta\right)$ is expressed as
\begin{equation}
  \frac{dg_h}{d\eta}=\frac{\left(\int_{A}^{B}\frac{\exp\left(\frac{\eta}{u}\right)}{u}du\right)^2-\int_{A}^{B}\exp\left(\frac{\eta}{u}\right)du\int_{A}^{B}\frac{\exp\left(\frac{\eta}{u}\right)}{u^2}du}{\left(\int_{A}^{B}\frac{\exp\left(\frac{\eta}{u}\right)}{u}du\right)^2}.
\end{equation}

{\noindent}Applying the Cauchy-Schwarz inequality in the inner product space of $L\left[A,B\right]$ elicits the inequality of
\begin{equation}
  \left(\int_{A}^{B}\frac{\exp\left(\frac{\eta}{u}\right)}{u}du\right)^2\leq\int_{A}^{B}\exp\left(\frac{\eta}{u}\right)du\int_{A}^{B}\frac{\exp\left(\frac{\eta}{u}\right)}{u^2}du.
\end{equation}

{\noindent}As a result, $dg_h/d\eta\leq 0$ indicates that $g_h\left(\eta\right)$ is a monotone decreasing function on $(-\infty,0]$.


\section{Proof of \textit{Proposition 2}}\label{sec:append_2}
We define auxiliary random variables as
\begin{equation}
  U\triangleq\sqrt{X\left(B-X\right)},\quad V\triangleq\sqrt{\frac{B-X}{X}}.
\end{equation}

{\noindent}Since the second-order momentums of $U$ and $V$ exist, the Cauchy-Schwarz inequality for random variables yields the inequality of
\begin{equation}\label{eq:cs_inequality_rv}
  \begin{split}
    &\left(\mathbb{E}\left[B-X\right]\right)^2=\left(\mathbb{E}\left[UV\right]\right)^2\leq\mathbb{E}\left[U^2\right]\mathbb{E}\left[V^2\right]\\
    &=\mathbb{E}\left[BX-X^2\right]\mathbb{E}\left[\frac{B}{X}-1\right].
  \end{split}
\end{equation}

Subsequently, an upper bound for $\sigma_X^2$ can be derived from~(\ref{eq:cs_inequality_rv}) as
\begin{equation}\label{eq:varx_upper_bound_1}
  \sigma_X^2\leq B\mathbb{E}\left[X\right]-\frac{\left(B-\mathbb{E}\left[X\right]\right)^2}{B\mathbb{E}\left[\frac{1}{X}\right]-1}-\left(\mathbb{E}\left[X\right]\right)^2.
\end{equation}

{\noindent}The equality in~(\ref{eq:varx_upper_bound_1}) holds when 
\begin{equation}
  U\overset{\text{a.s.}}{=}\iota V,\quad\exists\ \iota>0,
\end{equation}

{\noindent}i.e., $U$ equals to $\iota V$ almost surely (a.s.), which is equivalent to $X=B$ or $X=\sqrt{\iota}$. Therefore, if the right-hand side of~(\ref{eq:varx_upper_bound_1}) can be maximized by a Bernoulli distribution, then \textit{Proposition 2} holds.

To maximize the right-hand side of~(\ref{eq:varx_upper_bound_1}), an auxiliary function is defined as
\begin{equation}
  g_\nu\left(u,v\right)\triangleq Bu-\frac{\left(B-u\right)^2}{Bv-1}-u^2,
\end{equation}

{\noindent}where $u$ and $v$ correspond to $\mathbb{E}\left[X\right]$ and $\mathbb{E}\left[1/X\right]$, respectively. Besides, considering the constraints of
\begin{equation}
  \mathbb{E}\left[\frac{\left(X-A\right)\left(X-B\right)}{X}\right]\leq 0,\quad\mathbb{E}\left[\frac{1}{X}\right]\leq\varsigma,
\end{equation}

{\noindent}the optimization problem for $\sigma_X^2$ is formulated as
\begin{subequations}\label{eq:var_opt}
  \begin{align}
    &\text{(P3):} & &\mathop{\max}\limits_{u,v}\ & &\sigma_X^2=g_\nu\left(u,v\right),&\label{eq:var_opt:obj}\\
    & & &\quad\text{s.t.}\ & &u+ABv\leq A+B,&\label{eq:var_opt:maxvar_constr}\\
    & & &\ & &v\leq\varsigma.&\label{eq:var_opt:harmonic_constr}
  \end{align}
\end{subequations}

Since $g_\nu\left(u,v\right)$ is a negative-definite quadratic function of $u$, $u^*=\left(Bv+1\right)/2v$ maximizes $g_\nu\left(u,v\right)$ in the absence of constraint (\ref{eq:var_opt:maxvar_constr}). However, constraint (\ref{eq:var_opt:maxvar_constr}) elicits an upper bound for a feasible $u$, which indicates that
\begin{equation}
  g_\nu\left(u^*,v\right)=
  \begin{cases}
    \frac{B\left(Bv-1\right)}{4v},\quad\quad\quad\quad\frac{Bv+1}{2v}\leq A+B-ABv,\\
    AB\left(-ABv^2+\left(A+B\right)v-1\right),\ \text{otherwise}.
  \end{cases}
\end{equation}

{\noindent}Towards this end, three cases are considered to derive the global optimal solution to~(P3).

\subsubsection{Case III-1}
If $\varsigma\geq\left(1/A+1/B\right)/2$, $g_\nu\left(u^*,v\right)$ is maximized when $v=\left(1/A+1/B\right)/2$, which also subjects to the constraint $\left(Bv+1\right)/2v\geq A+B-ABv$. In this case, the optimal solution is expressed as
\begin{subequations}
  \begin{align}
    &\mathbb{E}\left[X\right]=\frac{A+B}{2},\\
    &\mathbb{E}\left[\frac{1}{X}\right]=\frac{1}{2}\left(\frac{1}{A}+\frac{1}{B}\right),\\
    &\sigma_X^2=\frac{\left(B-A\right)^2}{4}.
  \end{align}
\end{subequations}

\subsubsection{Case III-2}
If $1/2A\leq\varsigma\leq\left(1/A+1/B\right)/2$, $g_\nu\left(u^*,v\right)$ is maximized when $v=\varsigma$. Given the fact that $\left(Bv+1\right)/2v\geq A+B-ABv$, the optimal value $u^*$ is restricted by~(\ref{eq:var_opt:maxvar_constr}), which is equivalent to
\begin{subequations}
  \begin{align}
    &\mathbb{E}\left[X\right]=A+B-AB\varsigma,\\
    &\mathbb{E}\left[\frac{1}{X}\right]=\varsigma,\\
    &\sigma_X^2=\left(\varsigma-\frac{1}{B}\right)\left(\frac{1}{A}-\varsigma\right)A^2B^2.
  \end{align}
\end{subequations}

\subsubsection{Case III-3}
If $\varsigma\leq 1/2A$, $g_\nu\left(u^*,v\right)$ is also maximized when $v=\varsigma$. In this case, the inequality $1/B<1/2A$ holds due to the fact that $\varsigma\geq 1/B$. Since the inequality $\left(Bv+1\right)/2v\leq A+B-ABv$ holds when $v\in\left[1/B,1/2A\right]$, the optimal solution is given by
\begin{subequations}
  \begin{align}
    &\mathbb{E}\left[X\right]=\frac{B\varsigma+1}{2\varsigma},\\
    &\mathbb{E}\left[\frac{1}{X}\right]=\varsigma,\\
    &\sigma_X^2=\frac{B\left(B\varsigma-1\right)}{4\varsigma}.
  \end{align}
\end{subequations}

Since the maximization of $g_\nu\left(u,v\right)$ only involves the values of $\mathbb{E}\left[X\right]$ and $\mathbb{E}\left[1/X\right]$, a Bernoulli distribution can always be conceived under these two constraints, which completes the proof of \textit{Proposition 2}.

\begin{figure*}[!b]
  \normalsize
  \hrulefill
  \vspace*{4pt}
  \begin{equation}\label{eq:piece_wise_integral}
    \begin{split}
      \underbrace{\mathbb{E}\left[-\int_{-\infty}^{\infty}f_{Y\vert X}\left(y\right)\log\left(f_Y\left(y\right)\right)dy\right]}_{\mathcal{I}_0}=\underbrace{\mathbb{E}\left[\int_{-\infty}^{A-\delta}\frac{1}{\sqrt{2\pi}\tilde{\sigma}_c}\exp\left(-\frac{\left(y-X\right)^2}{2\tilde{\sigma}_c^2}\right)\log\left(\sqrt{2\pi}\tilde{\sigma}_c\exp\left(\frac{\left(y-A\right)^2}{2\tilde{\sigma}_c^2}\right)\right)\right]}_{\mathcal{I}_1}\\
      +\underbrace{\mathbb{E}\left[\int_{A-\delta}^{B+\delta}-\frac{\tilde{\eta}}{\sqrt{2\pi}\tilde{\sigma}_cy}\exp\left(-\frac{\left(y-X\right)^2}{2\tilde{\sigma}_c^2}\right)dy\right]}_{\mathcal{I}_{2,1}}+\underbrace{\mathbb{E}\left[\int_{A-\delta}^{B+\delta}\frac{\log\left(\mathcal{J}\left(\delta,\tilde{\eta}\right)\right)}{\sqrt{2\pi}\tilde{\sigma}_c}\exp\left(-\frac{\left(y-X\right)^2}{2\tilde{\sigma}_c^2}\right)dy\right]}_{\mathcal{I}_{2,2}}\\
      +\underbrace{\mathbb{E}\left[\int_{B+\delta}^{+\infty}\frac{1}{\sqrt{2\pi}\tilde{\sigma}_c}\exp\left(-\frac{\left(y-X\right)^2}{2\tilde{\sigma}_c^2}\right)\log\left(\sqrt{2\pi}\tilde{\sigma}_c\exp\left(\frac{\left(y-B\right)^2}{2\tilde{\sigma}_c^2}\right)\right)\right]}_{\mathcal{I}_3}.
    \end{split}
  \end{equation}
\end{figure*}

\section{Derivation of Upper Bound for High SNR}\label{sec:append_3}
A piece-wise integral with $f_Y\left(y\right)$ in~(\ref{eq:output_pdf_high}) elicits the expression of (\ref{eq:piece_wise_integral}) at the bottom of this page, where the integral in $\left[A-\delta,B+\delta\right]$ is split into two parts $\mathcal{I}_{2,1}$ and $\mathcal{I}_{2,2}$ for notational convenience. Subsequently, following the amplification techniques in~\cite{lapidoth_Capacity_FSO_2009}, the integral terms of Gaussian roll-off, i.e., $\mathcal{I}_1$ and $\mathcal{I}_3$ in~(\ref{eq:piece_wise_integral}), can be upper bounded by
\begin{subequations}
  \begin{align}
    &
    \begin{split}
      \mathcal{I}_1&\leq\frac{1}{2}Q\left(\frac{\delta}{\tilde{\sigma}_c}\right)+\frac{\delta}{2\sqrt{2\pi}\tilde{\sigma}_c}\exp\left(-\frac{\delta^2}{2\tilde{\sigma}_c^2}\right)\\
      &+\mathbb{E}\left[\log\left(\sqrt{2\pi}\tilde{\sigma}_c\right)Q\left(\frac{X-A+\delta}{\tilde{\sigma}_c}\right)\right],
    \end{split}
    \\
    &
    \begin{split}
      \mathcal{I}_3&\leq\frac{1}{2}Q\left(\frac{\delta}{\tilde{\sigma}_c}\right)+\frac{\delta}{2\sqrt{2\pi}\tilde{\sigma}_c}\exp\left(-\frac{\delta^2}{2\tilde{\sigma}_c^2}\right)\\
      &+\mathbb{E}\left[\log\left(\sqrt{2\pi}\tilde{\sigma}_c\right)Q\left(\frac{B-X+\delta}{\tilde{\sigma}_c}\right)\right].
    \end{split}
  \end{align}
\end{subequations}

Additionally, since $g_I\left(y\right)=1/y$ is a convex function of $y$ on $\left[A-\delta,B+\delta\right]$ with $A-\delta>0$, a piece-wise upper bound $\bar{g}_I\left(y\right)$ of $g_I\left(y\right)$ is given by
\begin{equation}
  g_I\left(y\right)\leq \bar{g}_I\left(y\right)=
  \begin{cases}
    -\dfrac{y-X}{\left(A-\delta\right)X}+\dfrac{1}{X},\ &A-\delta\leq y\leq X,\\
    -\dfrac{y-X}{\left(B+\delta\right)X}+\dfrac{1}{X},\ &X\leq y\leq B+\delta.
  \end{cases}
\end{equation}

Then, the integral term $\mathcal{I}_{2,1}$ in~(\ref{eq:piece_wise_integral}) can be upper-bounded in a piece-wise manner as
\begin{equation}\label{eq:integral_inv_piece}
  \begin{split}
    &\mathcal{I}_{2,1}\leq\mathbb{E}\left[\int_{A-\delta}^{B+\delta}-\frac{\tilde{\eta}\bar{g}_I\left(y\right)}{\sqrt{2\pi}\tilde{\sigma}_c}\exp\left(-\frac{\left(y-X\right)^2}{2\tilde{\sigma}_c^2}\right)dy\right]\\
    &=\mathbb{E}\Biggl[-\frac{\tilde{\eta}}{X}\Biggl(1-Q\left(\frac{X-A+\delta}{\tilde{\sigma}_c}\right)-Q\left(\frac{B-X+\delta}{\tilde{\sigma}_c}\right)\\
    &\quad+\frac{\tilde{\sigma}_c}{\sqrt{2\pi}}\Biggl(\frac{1}{A-\delta}\left(1-\exp\left(-\frac{\left(X-A+\delta\right)^2}{2\tilde{\sigma}_c^2}\right)\right)\\
    &\quad+\frac{1}{B+\delta}\left(\exp\left(-\frac{\left(B-X+\delta\right)^2}{2\tilde{\sigma}_c^2}\right)-1\right)\Biggr)\Biggr)\Biggr]\\
    &\overset{\text{(a)}}{\leq}\mathbb{E}\left[-\frac{\tilde{\eta}}{X}\right]\Biggl(1-2Q\left(\frac{B-A+2\delta}{2\tilde{\sigma}_c}\right)\\
    &\quad+\frac{\tilde{\sigma}_c}{\sqrt{2\pi}}\Biggl(\frac{1}{A-\delta}\left(1-\exp\left(-\frac{\left(B-A+\delta\right)^2}{2\tilde{\sigma}_c^2}\right)\right)\\
    &\quad-\frac{1}{B+\delta}\left(1-\exp\left(-\frac{\delta^2}{2\tilde{\sigma}_c^2}\right)\right)\Biggr)\Biggr),
  \end{split}
\end{equation}

{\noindent}where (\ref{eq:integral_inv_piece}a) follows the constraints of~(\ref{eq:capacity_opt:min_value_constr}) and (\ref{eq:capacity_opt:max_value_constr}).

Similarly, an upper bound for $\mathcal{I}_{2,2}$ under the constraint of~(\ref{eq:capacity_opt:harmonic_constr}) is expressed as
\begin{equation}
  \begin{split}
    &\mathcal{I}_{2,2}\leq\log\left(\mathcal{J}\left(\delta,\tilde{\eta}\right)\right)\\
    &\cdot\left(1-Q\left(\frac{1/\varsigma^*-A+\delta}{\tilde{\sigma}_c}\right)-Q\left(\frac{B-1/\varsigma^*+\delta}{\tilde{\sigma}_c}\right)\right),
  \end{split}
\end{equation}

{\noindent}where the value of $\varsigma^*$ is given by~(\ref{eq:varsigma_star}).

Consequently, assembling the upper bounds of $\mathcal{I}_1$, $\mathcal{I}_{2,1}$, $\mathcal{I}_{2,2}$, and $\mathcal{I}_3$ provides an upper bound for $\mathcal{I}_0$, which yields the high-SNR upper bound for channel capacity in~(\ref{eq:capacity_upper_bound_high}).

}


\bibliographystyle{IEEEtran}
\bibliography{Ref.bib}

\vfill

\end{document}